\documentclass[lettersize,journal]{IEEEtran}
\usepackage{amsmath,amsfonts}
\usepackage{array}
\usepackage{textcomp}
\usepackage{stfloats}
\usepackage{url}
\usepackage{verbatim}
\usepackage{graphicx}
\usepackage{orcidlink}
\usepackage{cite}
\usepackage{hyperref}
\usepackage{url} 

\usepackage{rotating}
\usepackage{algorithm} 
\usepackage{algpseudocode} 
\usepackage{amsmath,amsfonts,amssymb,amsthm,epsfig,epstopdf,array}

\theoremstyle{definition} 
\usepackage{mathtools}
\newtheorem{definition}{Definition}

\theoremstyle{plain}

\theoremstyle{remark}

\usepackage{lineno}
\usepackage{multirow}
\newcommand{\prob}{\mathbb{P}}
\newcommand{\cpt}{\mathbb{CPT}}

\usepackage{graphicx}
\usepackage{subfigure}

\begin{document}

\title{Risk Sensitivity in Markov Games and Multi-Agent Reinforcement Learning: A Systematic Review}

\author{Hafez Ghaemi~\orcidlink{0000-0001-6326-5258}, Shirin Jamshidi~\orcidlink{0009-0004-7642-9396}\textsuperscript{\textsection}, Mohammad Mashreghi~\orcidlink{0009-0008-0878-4926}\textsuperscript{\textsection}, Majid Nili Ahmadabadi~\orcidlink{0000-0002-6370-6057}, and Hamed Kebriaei~\orcidlink{0000-0002-3781-2163},~\textit{Senior Member,~IEEE}
\thanks{Correspondence: \textit{ghaemi.hafez@ut.ac.ir}.\\
\indent Hafez Ghaemi, Shirin Jamshidi, Mohammad Mashreghi, Majid Nili Ahmadabadi, and Hamed Kebriaei are with the School of Electrical and Computer Engineering, University of Tehran, Iran
}
}


\markboth{Preprint}%
{Shell \MakeLowercase{\textit{et al.}}: A Sample Article Using IEEEtran.cls for IEEE Journals}


\maketitle
\begingroup\renewcommand\thefootnote{\textsection}
\footnotetext{Equal contribution}
\endgroup
\begin{abstract}
Markov games (MGs) and multi-agent reinforcement learning (MARL) are studied to model decision making in multi-agent systems. Traditionally, the objective in MG and MARL has been risk-neutral, i.e., agents are assumed to optimize a performance metric such as expected return, without taking into account subjective or cognitive preferences of themselves or of other agents. However, ignoring such preferences leads to inaccurate models of decision making in many real-world scenarios in finance, operations research, and behavioral economics. Therefore, when these preferences are present, it is necessary to incorporate a suitable measure of risk into the optimization objective of agents, which opens the door to risk-sensitive MG and MARL. In this paper, we systemically review the literature on risk sensitivity in MG and MARL that has been growing in recent years alongside other areas of reinforcement learning and game theory. We define and mathematically describe different risk measures used in MG and MARL and individually for each measure, discuss articles that incorporate it. Finally, we identify recent trends in theoretical and applied works in the field and discuss possible directions of future research.
\end{abstract}


\section{Introduction}
\label{sec:intro}
\IEEEPARstart{M}{arkov} game (MG), also known as stochastic game (SG), is the main theoretical framework for studying multi-agent systems (MAS). This type of game was initially introduced by Shapley \cite{shapley1953stochastic} and was later formalized as the framework for multi-agent reinforcement learning (MARL) by Littman \cite{littman1994markov}. Within the classical MG and MARL paradigm, agents are assumed to pursue a risk-neutral objective, seeking to maximize a notion of expected return and ignore the subjective preferences of themselves or other agents within the MAS. Risk-neutral analysis of MGs and MARL has witnessed substantial progress in the recent years, particularly in specific types of MG such as stochastic zero-sum games \cite{sayin2021decentralized,zhang2020model,alacaoglu2022natural,perolat2015approximate,qiu2021provably,chen2024finite,park2024multi} and stochastic potential games \cite{ding2022independent,fox2022independent,mguni2021learning,leonardos2021global,maheshwari2022independent}. Nonetheless, this risk-neutral objective is often inadequate for representing agents with distinct subjective preferences and internal cognitive biases. To take such preferences into account, agents incorporate a measure of risk into their optimization objective, thus transitioning into the domain of risk-sensitive MG. To further elucidate why incorporating risk in MGs is important, we consider two practical scenarios and highlight the risk-sensitive nature of the objective in each.

\paragraph{Financial markets} The traditional objective of maximizing the \emph{expected} return is clearly not sufficient for many investors in financial markets. Most investors do not tolerate the possibility of extreme losses and would like to reduce the "risk" of their investments. Although one can try to model financial investment via single-agent risk-sensitive stochastic optimization and reinforcement learning (RL), the financial market is inherently an MAS, and a more accurate model of decision making for agents in this system would be risk-sensitive stochastic optimization in MGs and learning policies via risk-sensitive MARL.

\paragraph{Autonomous driving} The replacement of traditional cars with autonomous vehicles seems inevitable in the near future. A network of self-driving vehicles can be considered a complex MAS with objectives that can vary from agent to agent, for example, depending on subjective preferences of passengers such as the importance of punctuality, safety, and choice of itinerary. Again, in this scenario, the existence of multiple risk-sensitive agents with diverse preferences cannot be modeled without considering interactions between agents, and requires a multi-agent risk-sensitive framework.

Compared to single-agent stochastic optimization and RL, the literature on risk sensitivity in MG and MARL has been sparse. However, in recent years, researchers have realized the importance of incorporating risk into their multi-agent frameworks in areas such as finance, energy trade, operations research, behavioral economics, and cognitive modeling of human decision making. Although there exist many review and survey articles on MARL and MG \cite{zhang2021multi,canese2021multi,yang2020overview,busoniu2008comprehensive,nguyen2020deep,gronauer2022multi}, to the best of our knowledge, risk sensitivity in multi-agent decision making has not been explored in a review paper before. \textbf{To fill this gap, in this work, we present the first review paper on risk sensitivity in MG and MARL} by identifying the different risk measures used in MG and MARL, categorizing them, and discussing each measure's related articles in detail. The remainder of this review is organized as follows; Section~\ref{sec:pre} formally introduces and lays down the mathematical foundations of risk-sensitive MG and MARL and the different measures of risks used in these domains. In Section~\ref{sec:method}, we discuss the methodology of the review. Section~\ref{sec:res} is dedicated to analyzing and discussing the existing literature and selected articles. Finally, in Section~\ref{sec:discussion} we summarize our findings and identify the trends and future directions of research in risk-sensitive MG and MARL.

\section{Preliminaries}
\label{sec:pre}

A Markov game can be formally defined using a multi-player\footnote{We use the terms player and agent interchangeably in this paper. In general, player is more common in the game theory literature while agent is often used in RL and MARL literature.} Markov decision process (MDP) \footnote{Here, we define a discrete-time MDP. Similarly, continuous-time MDP and continuous-time MG can be defined. Some of the papers analyzed in this review work with continuous-time MGs.}.

\begin{definition}[Markov game] \label{def:pg}
    A stochastic (Markov) game\footnote{It is important to distinguish between Markov games where agents are assumed to make their decisions simultaneously and extensive-form games where agents take their actions sequentially after observing the actions of previous players.} is defined as a multi-player MDP using a tuple of five key elements $(N, \mathcal{S},\mathcal{A}=\{\mathcal{A}_i|i\in N\},\mathcal{P},R=\{R_i|i \in \mathcal{N}\})$, where,
\begin{itemize}
    \item $\mathcal{N}=\{1,2,3,...,n\}$ is a set of $n$ players.
    \item $\mathcal{S}$ is the state space shared by all player.
    \item $\mathcal{A}_i$ is the action space of player $i$.
    \item $\mathcal{P}:\mathcal{S}\times \mathcal{A}\rightarrow \Delta(\mathcal{S})$ is a transition probability mapping.
    \item $\mathcal{R}_i: \mathcal{S}\times \mathcal{A}\times\mathcal{S}\rightarrow \mathbb{R}$ is the reward function of player $i$.
\end{itemize}
\end{definition}

The joint policy of agents, probability distributions over actions in each state, is denoted by $\pi=(\pi^i, \pi^{-i})$. The framework for learning optimal policies in MGs is multi-agent reinforcement learning. Learning in MARL can be done in a cooperative, competitive, or mixed setting. Denoting the joint action profile at time $t$ by $a_t=(a^i_t, a^{-i}_t)$, the optimization objective for agent $i$ in the infinite-horizon discounted-reward MG is given by

\begin{equation} 
\small
    \max_{\pi^i} J_{\pi^i,\pi^{-i}}= \max_{\pi^i}\mathbb{E}_{a_t\sim \pi(.|s_t),s_{t+1}\sim p(.|s_t,a_t)}\bigg\{\sum_{t=0}^{\infty}\gamma^t r^i(s_t,a_t)\bigg\},
\end{equation}

where $\gamma$ is a discount factor. In the average-reward setting the objective is given by

\begin{equation} 
\tiny
    \max_{\pi^i} J_{\pi^i,\pi^{-i}} = \max_{\pi^i} \lim_{T\rightarrow \infty}\frac{1}{T} \mathbb{E}_{a_t\sim \pi(.|s_t),s_{t+1}\sim p(.|s_t,a_t)}\bigg\{\sum_{t=0}^{\infty} r^i(s_t,a_t)\bigg\}.
\end{equation}

Assuming that the policy $\pi^i$ is parameterized by variable $\theta$, we can  write the objective as

\begin{equation} \label{eq:noriskopt}
    \max_{\theta} J_{\pi^i,\pi^{-i}}(\theta).
\end{equation}

In general, in competitive settings, selfish agents reach a Markov perfect equilibrium (MPE) policy, also known as a Nash equilibrium (NE) of the MG, from which no agent can unilaterally deviate to increase its expected return without decreasing another agent's return. In the cooperative setting, agents would like to converge to a Pareto optimal policy in which there exists no alternative policy where at least one player's expected return is higher.

\subsection{Risk Measures in MG and MARL}\label{sec:riskm}

Risk-sensitive agents in stochastic optimization and RL, whether in the single-agent or multi-agent settings do not merely maximize an expected return objective, and deviate from it based on their specific measure of risk. Incorporating risk into an RL framework can be mathematically interpreted as modifying the underlying stochastic optimization. This modification is done by either changing the objective function, or by putting constraints on the optimization problem with the same objective function as the risk-neutral setting. Prashanth and Fu \cite{prashanth2022risk} refer the risk measures corresponding to the former as \emph{explicit} risk, and the ones corresponding to the latter as \emph{implicit} risk.

An explicit risk measure alters the objective function in MARL's stochastic optimization problem. Therefore, agent $i$ maximizes a function different from $J$ given in \eqref{eq:noriskopt}, such as

\begin{equation} \label{eq:expriskopt}
    \max_{\theta} G_{\pi^i,\pi^{-i}}(\theta).
\end{equation}

For implicit risk measures, the objective function remains the same as the risk-neutral one \eqref{eq:noriskopt}, but we will have a risk function $G$ and a risk level $\kappa$ that would convert the task into a constrained optimization,

\begin{equation} \label{eq:impriskopt}
    \max_{\theta} J_{\pi^i,\pi^{-i}}(\theta) \quad s.t. \quad G(\theta) < \kappa.
\end{equation}

Note that we can adopt methods used in constraint optimization such as Lagrange multipliers to transform an implicit risk into an explicit one, which is usually done in practice. In the following subsections, we will describe the risk measures in both explicit and implicit categories that we have identified in the literature of MG and MARL.

\subsection{Explicit Risk Measures}
\label{sec:exprisk}

There are three types of explicit risk measures used in MG and MARL; exponential reward, coherent risk, and cumulative prospect theory (CPT).

\subsubsection{Exponential Reward/Cost}

Exponential reward (or cost in the minimization case) is the most frequently used measure in risk-sensitive control, finance, and operations research. The classical formulation of exponential reward in risk-sensitive MDP was given by Howard and Matheson \cite{howard1972risk}. We can consider exponential reward in both discounted-reward and average-reward settings. The exponential reward objective in a discounted-reward MDP can be written as

\begin{equation} \label{eq:expcostdis}
    \max_{\pi^i} J_{\pi^i,\pi^{-i}} = \mathbb{E}_\pi [e^{\beta_i R_i}] ,
\end{equation}
where,
\begin{equation} \label{eq:expcostdis}
    \quad R_i = \sum_{t=0}^{\ T-1}\gamma_t r_{t,i},
\end{equation}
where $R_i$ is the cumulative reward of agent $i$ over an episode and $\beta$ is the hyperparameter that controls risk sensitivity; $\beta < 0$ corresponds to the risk-averse setting,  $\beta > 0$ to the risk-seeking settings and the limit $\beta\rightarrow 0$ reduces the objective to the risk-neutral case.

In an average-reward MDP, the risk-sensitive objective can be written as

\begin{equation} \label{eq:expcostavg}
    \max_{\pi^i} G_{\pi^i,\pi^{-i}} = \max_{\pi^i} \limsup_{T\rightarrow \infty}\frac{1}{T} \frac{1}{\beta}\log \mathbb{E}\bigg\{exp(\beta\sum_{t=0}^{\infty} r^i(s_t,a_t))\bigg\},
\end{equation}

where $\beta$ is again the risk sensitivity coefficient.

\subsubsection{Coherent Risk}
\label{sec:Coherent}
First introduced by Artzner et al. \cite{artzner1999coherent}, a coherent risk measure $\rho$ (equivalent to $G$ in \eqref{eq:expriskopt}) is an explicit measure of risk that possesses four mathematical properties ($X$ and $Y$ are random variables (r.v.s) that can represent returns of two different policies):

\begin{itemize}
  \item \textbf{Monotonicity:} If $X \leq Y$ (almost surely), then $\rho(X) \leq \rho(Y)$.
  \item \textbf{Sub-additivity}: $\rho(X+Y)\leq \rho(X) + \rho(Y)$.
  \item \textbf{Positive homogeneity}: $\rho(\lambda X) = \lambda \rho(X)$ for any $\lambda \geq 0$.
  \item \textbf{Translation invariance}: For constant $\alpha \geq 0$,  $\rho(X + \alpha) = \rho(X) + \alpha$.
\end{itemize}

In a risk-sensitive MDP, the r.v. $X$ can represent the cumulative discounted or average reward/cost. The above properties also have investment-related interpretations. For instance, sub-additivity is necessary in portfolio optimization implying that diversification is intended to mitigate the possibility of increased risk (in a cost minimization scenario). Another important property of this family of risk measures is that any coherent risk measure has a dual representation as the supremum of an expected value over a risk envelope \cite{rockafellar2007coherent}; i.e.,

\begin{equation}
  \mathcal{\rho}(X)=\sup _{Q \in \mathcal{Q}} \mathbb{E}(X Q).
\end{equation}

It follows that the risk envelope can be explicitly written as

\begin{equation}
    \mathcal{Q}=\left\{Q \in \mathcal{P}: \mathbb{E}(X Q) \leq \rho(X) \text { for all } X \in \mathcal{L}^2\right\},
\end{equation}

where $\mathcal{L}^2$ denotes $X$ satisfying $E(\lvert X \rvert^2)<\infty$. This "dual representation" is useful when one wants to prove a risk measure is coherent.

\subsubsection{Cumulative Prospect Theory}

Cumulative prospect theory (CPT) \cite{kahneman1979prospect,tversky1992advances} is a risk measure that models human attitudes toward risk when making decisions under uncertainty. It considers weighting functions that are applied to cumulative probabilities of outcomes to account for the fact that humans overestimate/underestimate outcomes with small/large probabilities. It also applies a usually convex-concave utility function to rewards to model humans' loss aversion and diminishing marginal utility. When considering CPT as an explicit risk measure, the expectation operator in \eqref{eq:expriskopt} is replaced by the non-linear CPT operator. We formally define the CPT operator and the corresponding risk-sensitive objective in MG and MARL below.

\begin{figure}[h]
 \centering
 \begin{subfigure}
  \centering
  \includegraphics[scale=0.3]{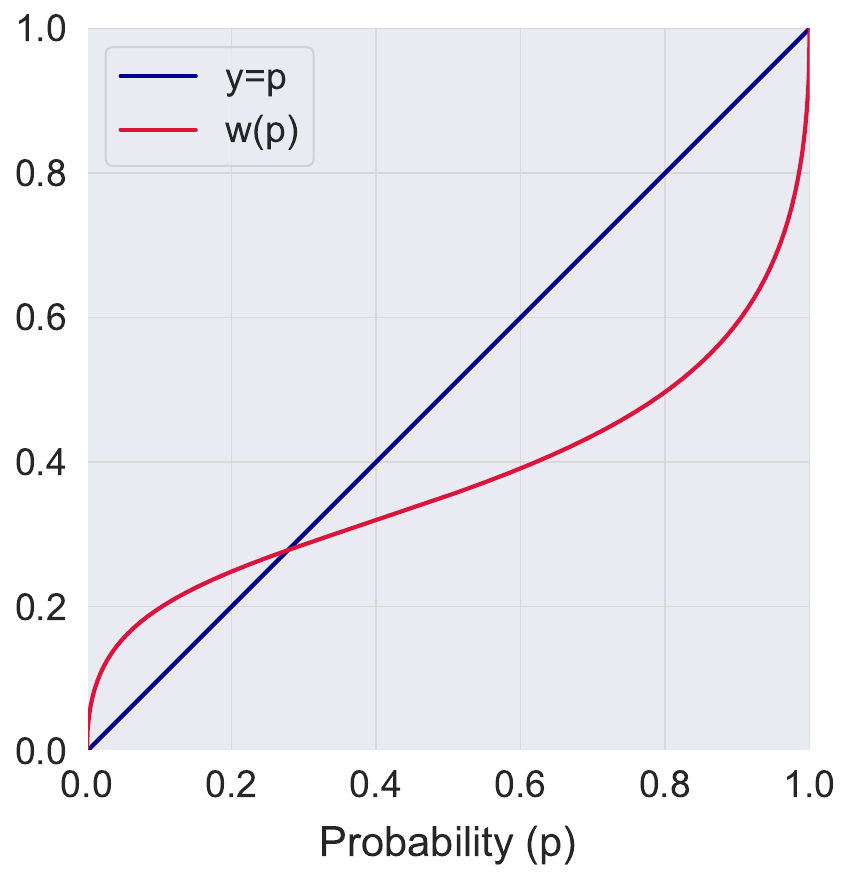}
  \label{fig:weighting}
 \end{subfigure}
 \hfill
 \begin{subfigure}
  \centering
  \includegraphics[scale=0.27]{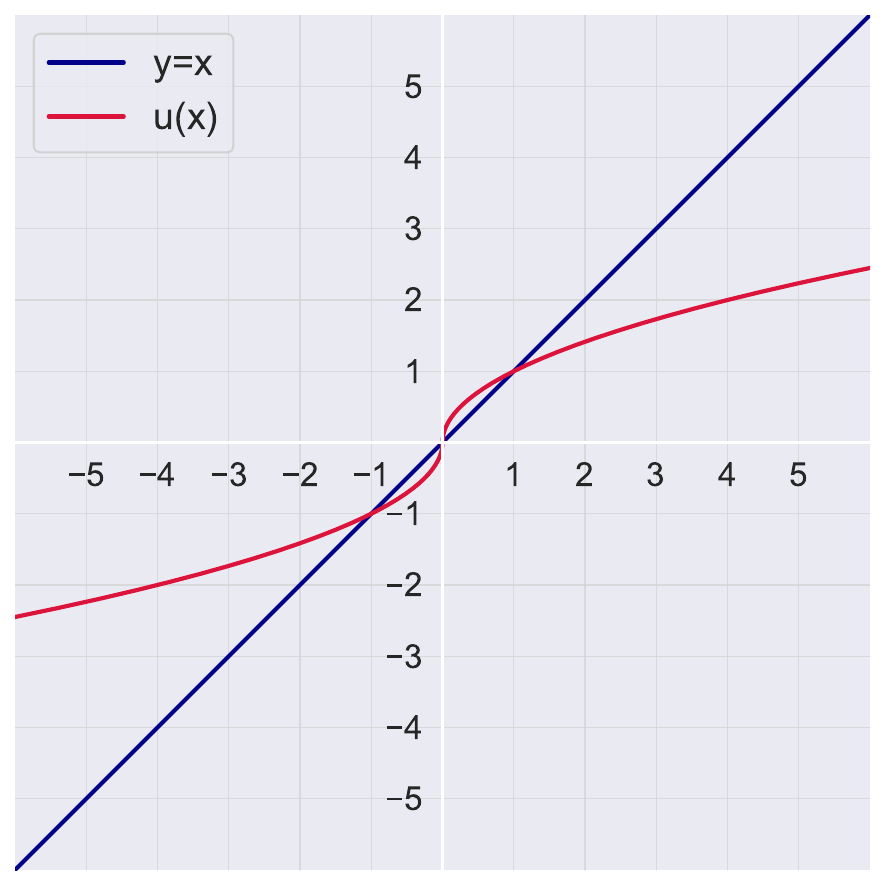}
  \label{fig:util}
 \end{subfigure}
 \caption{(Left) Conventional CPT weighting functions; $\omega^+(p) = \frac{p^{\gamma}}{(p^{\gamma} + (1-p)^{\gamma})^{(1/\gamma)}}$ and $\omega^-(p) = \frac{p^{\delta}}{(p^{\delta} + (1-p)^{\delta})^{(1/\delta)}}$ with $\gamma=\delta=0.69$. (Right) Conventional CPT utility functions; $u^+(x)=x^{\alpha}$ for $x\geq 0$, and $-u^-(x)=-\lambda(-x)^{\beta})$ for $x<0$, with $\alpha=\beta=0.65$ and $\lambda=2.6$.}
 \label{fig:combined}
\end{figure}

Given a real-valued r.v. $X$ with distribution $\prob(X)$, a reference point $x_0$, two monotonically non-decreasing weighting functions, $\omega^+: [0,1] \rightarrow [0,1], \omega^-: [0,1] \rightarrow [0,1]$, utility functions $u^+: \mathbb{R}^+ \rightarrow \mathbb{R}^+, u^-: \mathbb{R}^-\rightarrow \mathbb{R}^+$, and appropriate integrability assumptions, we can define the CPT value using Choquet integrals as

\begin{equation}
\begin{split}
    \cpt_{\prob}[X] &:= \int^{\infty}_0 \omega^+(\prob(u^+((X-x_0)_+)>x))dx-\\ &\int^{\infty}_0 \omega^-(\prob(u^-((X-x_0)_-)>x))dx., 
\end{split}
\end{equation}

where we denote $(.)_+=max(0,.)$ and $(.)_+=-min(0,.)$, and $x_0$ serves as a reference point that separates gains and losses. Without loss of generality, we can assume $x_0=0$. Conventional representations of CPT weighting functions are $\omega^+(p) = \frac{p^{\gamma}}{(p^{\gamma} + (1-p)^{\gamma})^{(1/\gamma)}}$ and $\omega^-(p) = \frac{p^{\delta}}{(p^{\delta} + (1-p)^{\delta})^{(1/\delta)}}$ \cite{tversky1992advances}. Note that by setting $\delta$ and $\gamma$ equal to $1$, the definition of expected utility $E_{\prob}[u(X)]$ is recovered which shows that CPT is a generalization of expected utility theory. Furthermore, $u^+$ and $u^-$ are usually concave functions ($-u^-$ is convex) to reflect the higher sensitivity of humans towards losses compared to gains, and the principle of diminishing marginal utility. Therefore, the utility function can have analytical representations $u^+(x)=x^{\alpha}$ if $x\geq 0$, and $u^-(x)=\lambda (-x)^{\beta}$ if $x<0$. The parameters $\gamma, \delta, \alpha, \beta,$ and $\lambda$ are subjective model parameters that can differ from person to person based on their level of risk aversion and individual characteristics. The conventional representations of weighting and utility functions given a set of subjective parameters are plotted in Figure~\ref{fig:combined}. It is noteworthy that CPT is a generalization of coherent measures of risk as it only possesses monotonicity and positive homogeneity but it is neither translation-invariant nor convex. Indeed, by choosing appropriate CPT weighting functions, one can derive different coherent risk measure formulations \cite{jie2018stochastic}. It is also noteworthy that since CPT can be seen as a weighted risk measure with a utility function and distorted probabilities, its different variants that are similarly defined using the aforementioned Choquet integral, have been called distortion risk measure (DRM) \cite{denneberg1990distorted,sereda2010distortion,vijayan2021policy}. Due to their similar mathematical definition, we do not distinguish between CPT and DRM in this paper.

\subsection{Implicit Risk Measures}
\label{sec:imprisk}

There are three types of implicit risk measures used in MG and MARL; variance, conditional value-at-risk, and chance constraint.

\subsubsection{Variance}
When using variance as a risk measure, we are interested in imposing a constraint on the variance of returns. In discounted MDPs, variance as a risk was first introduced by Sobel \cite{sobel1982variance}. In this setting, we would like to upper-bound the \emph{overall} variance of cumulative discounted return. Therefore, the risk constraint $G$ in \eqref{eq:impriskopt} will be

\begin{equation} \label{eq:varcostdistG}
\small
    G_{\pi^i,\pi^{-i}}(s_{0}) = Var[D_{\pi^i,\pi^{-i}}(s_{0})]= U_{\pi^i,\pi^{-i}}(s_{0})-J_{\pi^i,\pi^{-i}}(s_{0})^{2},
\end{equation}

where,

\begin{equation} \label{eq:varcostdistJ}
\small
    J_{\pi^i,\pi^{-i}}(s) = \mathbb{E}[D_{\pi^i,\pi^{-i}}(s)] \quad and \quad U_{\pi^i,\pi^{-i}}(s) = \mathbb{E}[D_{\pi^i,\pi^{-i}}(s)^{2}],
\end{equation}

\begin{equation} \label{eq:varcostdistJ}
    D_{i}(s_{0}) = \sum_{t=0}^{\infty}\gamma^t r^i(s_t, a_t),
\end{equation}

Where $D_i$ represents the total discounted reward for a trajectory in an infinite-horizon MDP.

On the other hand, in average-reward MDPs, Filar et al. \cite{filar1989variance} introduced the per-period variance defined by deviations of single-stage reward from the average reward, in contrast to the variance of average reward itself. Therefore, the constraint in \eqref{eq:impriskopt} will be

\begin{equation} \label{eq:varcostdistG_avg}
    G_{\pi^i,\pi^{-i}} = \lim_{T\rightarrow \infty}\frac{1}{T}\mathbb{E} \bigg\{\sum_{t=0}^{\ T-1} (r^i(s_t,a_t)-J_{\pi^i,\pi^{-i}})^2\bigg\}.
\end{equation}

An example can show the rationale behind the choice of this constraint instead of the variance of average reward itself. Consider two sequences of rewards; $(20, -20, 20, -20, ...)$ resulting from policy $\pi_1$ and $(0,0,0,0, ...)$ resulting from $\pi_2$. The average reward and its variance are zero for both $\pi_1$ and $\pi_2$, however, it is evident that in terms of the constraint \eqref{eq:varcostdistG_avg}, policy $\pi_2$ is better than $\pi_1$.

\subsubsection{Conditional Value-at-Risk}

The Value-at-risk (VaR) is a risk measure widely employed in financial applications. For an r.v. $X$, VaR at a given level $\alpha \in (0,1)$ is defined as 

\begin{equation} \label{eq:varcostdistG}
    VaR_\alpha(X) \triangleq \inf \lbrace \emph{z} | \mathbb{P} ( X \leqslant \emph{z}) \geqslant \alpha 	 \rbrace	 	,
\end{equation}

where $\alpha \in [0,1)$ is an $\alpha$-quantile, and $F$ is the cumulative distribution function (c.d.f) of $X$. Therefore, the definition of VaR is equivalent to

\begin{equation} \label{eq:varcostdistG}
    VaR_\alpha(X) \triangleq \inf \lbrace \emph{z} | F_X ( \emph{z} ) \geqslant \alpha 	 \rbrace = F_X^{-1} (\alpha).
\end{equation}

It can be seen that VaR quantifies the level of assets required to cover a potential loss. The main drawback of VaR is that it is not a coherent risk measure (in order to check if an implicit risk measure is coherent, we can use the dual representation theorem discussed in Section~\ref{sec:Coherent}). The mathematical properties of coherent risks make them compatible with standard stochastic optimization methods. In order to take advantage of such properties, we can modify VaR to derive another risk measure which is coherent, known as conditional value-at-risk (CVaR), first introduced by Rockafellar and Uryasev \cite{rockafellar2000optimization}. Considering the tail distribution described by VaR, CVaR is the conditional mean over it and is defined as

\begin{equation}
    \operatorname{CVaR}_\alpha(X) \triangleq \mathbb{E}\left[X \mid X \geq \operatorname{VaR}_\alpha(X)\right].
\end{equation}

Therefore, the corresponding risk constraint in \eqref{eq:expriskopt} is $X \geq \operatorname{VaR}_\alpha(X)$. It can be shown that CVaR has the dual representation of coherent risk measures (see \cite{ang2018dual} for detailed derivations), i.e.,

\begin{equation}
    \operatorname{CVaR}_\alpha(X)=\sup _{Q \in \mathcal{Q}_\alpha} \mathbb{E}(X Q)
\end{equation}

with

\begin{equation}
\mathcal{Q}_\alpha:=\left\{Q \in \mathcal{L}^2: \mathbb{E}(Q)=1,0 \leq Q \leq \frac{1}{1-\alpha}\right\}.
\end{equation}

Consequently, CVaR can also be considered an explicit measure of risk as it is coherent.

\subsubsection{Chance Constraint}

Chance constraints in stochastic optimization was first used by Charnes et al. \cite{charnes1958cost}. When using a chance constraint, we would like the probability that our incurred loss is larger than a specific amount to be smaller than a given threshold, i.e.,

\begin{equation} \label{eq:chanceconst}
    \mathbb{P}(g(X)\ge 0) \le \alpha  ,
\end{equation} 

where $g$ is a loss function defined with relation to return, $X$ is an r.v, and $\alpha$ is a small probability quantile. We can therefore write the constraint in the form of \eqref{eq:impriskopt} as $G \triangleq \mathbb{P}(g(X) \geq 0)$.

\section{Review Methodology}
\label{sec:method}

In this section, we outline the methodology used to conduct this review. Following the PRISMA guidelines for systematic reviews \cite{page2021prisma}, we conducted a keyword search to first extract all relevant articles. The final search was performed on May 24, 2024 to retrieve the up-to-date corpus of articles on Scopus database. We used the following query among title, abstract, and keywords of documents:

("Multi-Agent Reinforcement Learning" OR "Markov game" OR "Markov game" OR "Dynamic Games") AND ("Risk Sensitiv*" OR "Risk Measure*" OR "Cumulative Prospect Theory" OR "Conditional Value-at-Risk" OR "Chance Constraint" OR "Exponential Reward" OR "Exponential Cost").

This query without any additional filters gave 129 documents, among which there were 83 journal articles, 41 conference papers, 3 book chapters, 1 book, and no review article. All documents were in English, except for one that was in Chinese. We limited the articles to the English language and selected only journal articles and conference papers. The remaining number of documents was 124 after this stage. We further performed a manual search on Google Scholar and IEEE Xplore and found two additional recently published papers that have not been indexed by Scopus yet, resulting in a total of 126 articles. It is noteworthy that neither using the query search nor the additional manual search, we did not find a relevant review article, establishing the present work as the first review on risk-sensitive MG and MARL. Interestingly, even the area of single-agent risk-sensitive RL was lacking a comprehensive survey until very recently \cite{prashanth2022risk}.

We did not perform any additional pre-screening on the 126 documents and assessed all articles in full. During the full-text assessment, we excluded any article that did not meet the following criteria:

\begin{itemize}
    \item The underlying framework to be analyzed must be a stochastic MAS involving at least two or more agents. The MAS should be represented either explicitly or implicitly as a multi-agent MDP, where each agent optimizes an objective function in a cooperative, competitive or mixed manner.
    \item The objective must be risk-sensitive, i.e., at least one of the agents must deviate from the risk-neutral objective and incorporate a risk measure, i.e., the ones described in Section~\ref{sec:riskm}.
\end{itemize}

After the full-text assessment, we excluded 67 documents that did not meet the above eligibility criteria for risk-sensitive MG and MARL. Many of the excluded articles studied the relationship between zero-sum Markov games and single-agent risk-sensitive control \cite{Başar1999479,Golui2023327,zhang2021provably,Tang20031065,Dvijotham2011179,zhang2021derivative,Dey19971587,Fleming19971790,Ding202239}. For this reason, the search query used above could not exclude them automatically, and a full-text assessment was necessary to detect these studies and filter them. Finally, 59 remaining documents are included in our final analysis; we categorize these articles based on the risk measure used and discuss them individually in the following section.

\section{Risk Sensitivity in MG and MARL}
\label{sec:res}

In this section, we discuss the 59 studies selected using our selection procedure laid out in Section~\ref{sec:method}. We dedicate a subsection to each risk measure and discuss the papers related to each measure by considering the type of MG, the type of state and action spaces, the goal and result of the study, and the algorithms used.

\subsection{Explicit Risk Measures}

\subsubsection{Exponential Reward/Cost}

The exponential reward is the most frequently used risk measure in risk-sensitive MG and MARL. We separately consider the average-reward and the discounted-reward settings for exponential risk in this section.

\emph{1.1) Average Exponential Reward/Cost:} There are four studies that work with denumerable state spaces and Borel action spaces in discrete-time MGs and average-reward exponential risk \cite{Ghosh202340, Wei2021835, Wei2019risk,Hernández-Hernández1996147}. Ghosh et al. \cite{Ghosh202340} work in two-player zero-sum MGs and consider unbounded payoff functions. Via Lyapunov stability assumptions, the authors show the existence of the value of the game and a saddle point equilibrium and further characterize the corresponding equilibrium's stationary Markov policies analytically. Wei and Chen \cite{Wei2021835} focus on nonzero-sum two-player MGs. They first establish a Feynman–Kac formula for unbounded payoff functions under randomized history-dependent strategies, and afterwards prove the existence of solutions to optimality equations for both players and establish the presence of randomized stationary NE under specific conditions. In another work, Wei and Chen \cite{Wei2019risk} consider nonzero-sum coordination MGs with multiple players. They show the existence of solutions to coupled player optimality equations and confirm the presence of stationary NE under suitable conditions through a discounted approximation method. They also show that these results hold even with negative risk sensitivity coefficients. Lastly, Hernàndez-Hernàndez and Marcus \cite{Hernández-Hernández1996147} consider two-player competitive MGs with infinite horizons and bounded cost functions. They demonstrate the presence of a bounded solution to the dynamic programming value function equations by showing that they satisfy an Isaacs equation. Consequently, they prove the existence of optimal stationary policies for each player.

Considering continuous-time MGs with denumerable state spaces and Borel action spaces, Golui and Pal \cite{Golui202278} analyze zero-sum two-player MGs with a finite time horizon. Through a non-homogeneous game model, they establish the existence of the game value and saddle-point equilibrium with history-dependent policies. Wei \cite{Wei201996} studies multi-player nonzero-sum MGs with finite horizons and unbounded cost and transition rates. The paper proves the existence of a randomized MPE with history-dependent policies. Lastly, Ghosh et al. \cite{ghosh2022nonzero} examine two-player nonzero-sum MGs with infinite time horizons and unbounded cost and transition rates. Assuming a Lyapunov stability assumption and using the principal eigenvalue approach, they show that Hamilton–Jacobi–Bellman (HJB) equations admit a solution which implies the existence of NE and corresponding stationary Nash policies.

Shifting attention to MGs with continuous state and action spaces, Caravani and Papavassilopoulos \cite{Caravani1990117} investigate non-cooperative linear-quadratic stochastic dynamic games with conflicting risk aversions and incomplete information and establish conditions for the existence of a risk-sensitive NE. Wang et al. \cite{Wang20206998} consider stochastic dynamic games and propose an iterative algorithm that finds the feedback NE of the game by a linearized approximation of the system dynamics and a quadratic approximation of the cost function. They experimentally show that the risk-sensitive Nash equilibria result in safer driving behavior of agents in multiple driving scenarios. Lastly, Krajewski \cite{Krajewski1991260} delves in competitive linear exponential-quadratic stochastic dynamic games in one-step delay information-sharing patterns. The article shows that a unique NE exists in such a setting and can be computed recursively.

Two studies analyze inter-generational (bequest) MGs involving multiple players, in which the utility of each generation (time step) depends on actions of more than one descendant player \cite{Jaśkiewicz2014411,Balbus2015247}. Jaśkiewicz and Nowak \cite{Jaśkiewicz2014411} consider Borel state and action spaces. They combine risk-sensitive control theory with overlapping generations models to prove the existence of a stationary MPE in these games using the Dvoretzky, Wald, and Wolfowitz theorem. Balbus and Jaśkiewicz \cite{Balbus2015247} consider continuous state and action spaces, as well as weakly continuous transition probabilities that include both non-atomic and deterministic scenarios. They prove the existence of a stationary MPE using Fatou's lemma and Skorohod's representation theorem.

Klompstra \cite{Klompstra2000nash,Klompstra19952458} analyzes continous-time linear-quadratic nonzero-sum stochastic dynamic games with two players and continuous state and action spaces. These studies analyze both linear-exponential-Gaussian and linear-exponential-quadratic-Gaussian control problems to establish NE solutions both under complete state observation and shared partial observation. 

Xu and Wu \cite{Xu2023} explore stochastic dynamic games with continuous state and action spaces and a large population of minor agent and a major agent that has a significant influence on the minor agent in a linear-quadratic-Gaussian system. The authors establish a decentralized $\epsilon$-Nash equilibrium using the Nash certainty equivalence methodology and limiting control problem.

Naderi Soorki et al. \cite{NaderiSoorki20217444} model joint beam forming and phase shift control of millimeter wave communications via an identical-payoff partially observable MG with continuous state and action spaces. They design a recurrent neural network for each agent to find risk-sensitive optimal policies via distributed policy gradient (PG) search in the policy space and prove the convergence of their PG algorithm to the NE of the game.

Başar \cite{bacsar2021robust} studies different scenarios of risk-sensitive control within continuous-time MGs with continuous state and action spaces. After discussing the equivalence between single-agent risk-sensitive control and filter design with two-player zero-sum stochastic differential games, the paper also discusses equivalences between risk-sensitive two-player zero-sum stochastic differential games and risk-neutral three-player zero-sum stochastic differential games. The paper also discusses robustness issues in risk-sensitive multi-player mean-field stochastic differential games with finite or infinite number of players.

Bhabak and Saha \cite{Bhabak2023134} consider zero and nonzero-sum two-player semi-Markov non-homogeneous games with discrete-finite state and Borel action spaces, allowing for unbounded transition and cost rates under history-dependent strategies. They show the existence of the game value and saddle-point equilibrium. Via the Feynman-Kac formula, they also prove this equilibrium is unique. 

Ghosh et al. \cite{Ghosh2024risk} analyze two-player continuous-time zero-sum MGs with countable state and Borel action spaces. They prove the existence of the game value and saddle-point equilibrium under the Lyapunov stability conditions by establishing the existence of a principal eigenpair for the MG's Hamilton-Jacobi-Isaacs (HJI) equations. 

Lastly, Baier et al. \cite{baier2023} study exponential risk for the first time in turn-based minimax-style MGs with the total (average) reward objective and continuous state and action spaces. The authors show that these games are determined and admit optimal memoryless deterministic strategies, in contrast to games with other risk measures (such as variance or CVaR) that require randomization and memory. They provide results on the decidability and computational complexity of the threshold problem of whether the optimal exponential risk value exceeds a given threshold, showing that it is decidable subject to Shanuel's conjecture in the general case, but more tractable for rational inputs or small algebraic instances. The paper also provides an approximation algorithm for the optimal exponential risk value. 

\emph{1.2) Discounted Exponential Reward/Cost:} We start this subsection by studies on competitive zero-sum MGs. Three studies \cite{Golui2022485, Golui202278, Pal202213} focus on zero-sum continuous-time MGs with two players. Golui et al. \cite{Golui2022485} and Golui and Pal \cite{Golui202278} consider discrete state spaces and Borel action spaces and allow for unbounded transition and cost rates. They prove the existence of the game value and saddle-point equilibrium in the class of history-dependent strategies by imposing a Foster-Lyapunov condition and leveraging HJI equations. These works are a generalization of Pal and Pradhan \cite{Pal202213}, where the transition and cost rates are assumed to be bounded. Considering discrete-time zero-sum MGs, Guo et al. \cite{Guo2023} study two-player infinite-horizon case with Borel state and action spaces, unbounded payoffs, and an interesting condition where the discount factor varies as a function of the state. Considering Borel state and action spaces, they first find new boundaries for the Shapely equation (SE) under the unbounded payoffs condition, and afterwards, prove the existence of the solution to the SE and the existence of NE and value of the game. They also provide an iterative algorithm to find the NE of the game.

There are also studies that have considered competitive MGs that are not zero-sum. Asienkiewicz and Balbus \cite{Asienkiewicz2019502} focus on two-player resource extraction games with continuous state and action spaces, where agents are assumed to have identical preferences. They prove the existence of a symmetric stationary MPE given felicity utility functions for agents in the infinite-horizon setting. Monahan and Sobel \cite{Monahan1997149} consider a competitive setting between advertising firms modeled as a risk-sensitive stochastic dynamic game with continuous state and action spaces. The authors introduce the notion of a firm's "goodwill", i.e., the cumulative measure of the firm and its competitors' advertising expenditure and show that in a symmetric stochastic dynamic game with discounted-cost exponential risk, the equilibrium goodwill levels are inversely proportional to risk sensitivity, which has the consequence of lower initial advertising expenditure when risk sensitivity is high. Lastly, Wei and Chen \cite{wei2022risk} consider multi-player nonzero-sum MGs with countable state spaces and Borel action spaces. They prove the existence of the optimal value functions for each player and a randomized MPE.

Regarding coordination-based MARL in MGs, Noorani and Baras \cite{Noorani20222266} model a repeated two-player coordination stag-hunt game as an MG with discrete state and action spaces. They used REINFORCE PG algorithm \cite{williams1992simple} and experimentally showed that using a discounted exponential risk criterion would encourage coordination between agents and leads them to a Pareto optimal policy compared to a sub-optimal Nash equilibrium. 

\emph{1.3) Average and Discounted Exponential Reward/Cost:} The articles in this subsection study both discounted and average exponential risk.

Two studies \cite{Basu2018516,basu2014zero} focus on two-player games with countable state and continuous action spaces. Basu and Ghosh \cite{Basu2018516} analyze nonzero-sum MGs with infinite horizons and prove the existence of risk-sensitive NE policies for discounted cost under fairly general conditions, and for the average cost under a geometric ergodicity and smallness of payoff conditions on the MDP. On the other hand, Basu and Ghosh \cite{basu2014zero} analyze zero-sum MGs with infinite horizon and countable state and continuous action spaces. They prove the existence of game values and saddle-point NE of the game. They also provide upper and lower bounds for game values in both the discounted and average-cost cases. In the average-cost case, a uniform ergodicity condition is required for the results to hold. Finally, they show the relationship between the game value and the level of risk sensitivity in the average-cost case; they show that the game value is the product of the inverse of the risk sensitivity factor and the logarithm of the common Perron–Frobenius eigenvalue of the associated controlled nonlinear kernels.

Bäuerle and Rieder \cite{bauerle2022distributionally} study risk-sensitive zero-sum MGs in both discounted and average-reward cases with Borel state and action spaces and bounded rewards, with both finite and infinite time horizons. For the discounted-reward case, they prove the existence of the game value, which solves the Shapley equation, and the existence of the optimal possibly non-stationary strategies under continuity and compactness conditions. In the average-reward scenario, they show that game's value solves the Poisson's equation under a local minorization property and a Lyapunov condition. They also answer an open question posed by Basu and Ghosh \cite{basu2014zero} and prove the existence of optimal risk-sensitive policies for agents in the average-reward case.

Ghosh et al. \cite{Ghosh2016835} examine continuous-time MGs with a denumerable state and continuous action space, and prove the existence of the value and saddle-point equilibria under general conditions in the discounted-reward case, and in the average-reward case under certain Lyapunov conditions using the HJI equations.

Cavazos-Cadena and Hernández-Hernández \cite{Cavazos-Cadena2019219} consider a discrete-time MG with discrete state and compact metric action spaces. Under standard continuity-compactness assumptions and when the state space is communicating under deterministic stationary policies and a minorization property holds at some states, they show that as the discount factor $\alpha$ increases to 1, an appropriate normalization of the risk-sensitive discounted value function converges to the risk-sensitive average value function, characterized by the solution to the Shapley equation.

Lastly, Tembine \cite{Tembine20114264} introduces a mean-field MG involving multiple players, considering both discounted and average exponential risk with a finite-discrete state space and a one-dimensional compact action space. They derive backward-forward mean-field equations in the total payoff case. They also show that optimal policies for players may not exist in the average-reward case if the underlying MDP does not have a unique positive recurrent class.

\subsubsection{Coherent risk measures}
 
Frédéric Bonnans et al. \cite{FrédéricBonnans2021} investigate a discrete-time mean-field MG with coherent risk and continuous state and action spaces, where each agent controls a linear dynamical system. The model takes the form of a coupled system of dynamic programming equations and a Kolmogorov equation. The authors prove the existence of a solution to the coupled system obtained using Schauder's fixed point theorem. Finally, using the corresponding optimal feedback control from the mean-field game solution, they construct an approximate NE for the N-player MG with a quantified approximation error.

Pang et al. \cite{Pang2017235} study a non-cooperative MG with multiple players, where each agent solves a rival-parameterized stochastic program with quadratic recourse in a two-stage setting with continuous state and action spaces. In the first stage, agents take deterministic actions, and after the realized uncertainty, in the second stage they take recourse decisions. Each agent's objective function consists of a deterministic risk-neutral first-stage component and a second-stage coherent risk-sensitive component. They consider linear-quadratic recourse functions, and show that when using such functions, the agents' objectives are convex with respect to their decision variables if the first-stage components are convex. They propose smoothing and regularization techniques to obtain differentiable approximations of the total objective function and overcome the non-differentiability of the recourse functions. They also prove convergence of their best-response schemes to the NE of the MG.

Finally, Huang et al. \cite{Huang20202022} consider a non-cooperative MG with finite-discrete state and action spaces and coherent risk. They propose a risk-aware centralized Nash Q-learning that uses stochastic approximation for saddle-point problems to approximate the risk function. Under mild conditions, the authors prove the convergence of their algorithm to a risk-sensitive MPE of the MG.

\subsubsection{Cumulative Prospect Theory}

Tian et al. \cite{Tian20216011} use the nested CPT formulation in MDPs \cite{lin2018probabilistically} and bounded-rational agents with quantal level-$k$ policies \cite{wright2017predicting} and developed a framework called bounded risk-sensitive Markov game (BRSMG). Considering only finite-discrete state and action spaces and deterministic policies, they proposed a model-based value iteration algorithm to find optimal CPT-sensitive policies. The authors also provide an inverse reward learning algorithm to find the agents' reward functions, intelligence level, and CPT parameters in a two-player model-based scenario.

Xiao et al. \cite{Xiao20182512} apply a non-nested version of the dynamic CPT risk \cite{jie2018stochastic} in an MG with finite-discrete state and action spaces to study the interactions between a cyber system and an advanced persistent threat (APT) attacker when they make subjective decisions to choose their scan and attack intervals, respectively (actions in the risk-sensitive MG). The utility in this setting is a function of the data protection level and the scan interval. They first consider a single-state risk-sensitive game where the whole duration of the attack is a single-state game. Afterwards, they consider an MG where the dynamics of the APT model is not known by the agents and the current state is considered as a discretized safe duration of the previous time interval in the attack. The authors compare normal Q-learning and a modified policy hill-climbing (PHC) detection scheme, in which they utilize a hotbooting technique that uses experiences from similar scenarios to initialize the Q functions and accelerate learning speed of the PHC-based algorithm. The empirical results show that this modified PHC-based algorithm outperforms Q-learning in terms of the average CPT utility and data protection level.

Shen et al. \cite{shen2023riskq} consider the non-nested version of the dynamic CPT risk (which they refer to as distortion risk measure also encompassing VaR and CVaR) in a cooperative partially observable MG with finite-discrete state and action spaces. The authors introduced Risk-sensitive Individual-Global-Max (RIGM) principle for cooperative risk-sensitive MARL which ensures that the optimal joint risk-sensitive action is equivalent to the collection of each agent's greedy risk-sensitive actions. They then propose RiskQ, a decentralized algorithm based on implicit qunatile network (IQN) \cite{dabney2018implicit} that models the joint return distribution of agents by combining per-agent return distribution utilities using an attention-based mechanism to satisfy the RIGM principle. They also conduct experiments on multiple cooperative MARL environments showing that RiskQ outperforms existing methods in several scenarios.

Finally, Ghaemi et al. \cite{ghaemi2024risk} consider network aggregative Markov games (NAMG), a class of MGs with local communication between agents and reward functions that are dependent on neighboring agents \cite{moghaddam2024expected,parise2020distributed,shokri2020leader}. The authors propose an actor-critic style algorithm for distributed risk-sensitive MARL with a nested CPT risk criterion in NAMGs with finite-discrete state and action spaces. Under a set of assumptions, they prove the convergence of their algorithm to a risk-sensitive and subjective (from the agent's prespective) MPE. Experimentally, they show that higher loss aversion in CPT-sensitive agents leads to a higher tendency for social isolation in the MG. 

\subsection{Implicit Risk Measures}

\subsubsection{Variance}

We identified two studies using Variance as a risk, both of which operate in the discounted-reward setting.

Reddy et al. \cite{Reddy20192171} consider variance of return (VOR) as the risk measure in a multi-player mixed cooperative/competitive MARL scenario with continuous state and action spaces. They incorporate VOR as a constraint into the agent's policy optimization, transformed into an unconstrained problem using Lagrangian relaxation. The authors propose a multi-timescale actor-critic algorithm called RC-MADDPG for learning risk-constrained policies in MGs. The algorithm uses a centralized training, decentralized execution framework. Experiments on the Keep Away task \cite{lowe2017multi} demonstrate that the proposed method learns risk-averse policies that satisfy the VOR constraint while achieving rewards comparable to risk-neutral policies.

Parilina and Akimochkin \cite{Parilina20211} construct a risk-sensitive cooperative discrete-time MG framework with discrete state and action spaces. The framework incorporates mean-variance preferences as risk into the discounted-reward optimization problem. The authors use a max-min optimization approach to define the characteristic functions of sub-games. They derive the core of the cooperative MG by applying the method proposed by Sujis et al. \cite{suijs1999cooperative} to a mean-variance risk-sensitive setting.

\subsubsection{Conditional Value-at-Risk}

After exponential reward/cost, CVaR is the second most frequently used risk measure in MG and MARL. While the majority of works with exponential reward were theoretical, in case of CVaR, they are mostly applied.

He et al. \cite{He20211398} develop a CVaR-sensitive Cournot MG as an energy bidding framework for multiple wind farms with continuous state and action spaces. They propose an iterative algorithm to find the optimal risk-sensitive bid for each wind farm and validate its feasibility by numerical case studies.

Cui et al. \cite{Cui2020828} introduce a risk-sensitive CVaR energy sharing model for community photovoltaic systems as a competitive MG with continuous state and action spaces among photovoltaic prosumers using dynamic pricing. They utilize a relaxation method to find the sample weighted average approximation CVaR-sensitive NE of the game, where the weights correspond to different energy scenarios.

Li et al. \cite{Li20181386} propose a competitive CVaR-sensitive MG model with continuous state and action spaces to optimize plug-in electric vehicle (PEV) charging strategies within a smart grid. The CVaR risk is applied to the cost function to reduce the overload of the electric transformer. They prove the existence of a sampling average approximation (SAA) generalized NE for this game model and propose an iterative algorithm to find this NE using the Nikaido-Isoda relaxation method, in which the best response of each agent is calculated using a distributed alternating direction method of multipliers (ADMM) algorithm \cite{wei2012distributed}. They also provide numerical simulations using real-word data from a PEV taxi network.

Li et al. \cite{li2017risk} propose a CVaR-sensitive two-stage MG with continuous state and action spaces to quantify the overbidding risk for multi-energy microgrids, considering the risk from uncertain energy supply and demand. They use the Cournot Nash pricing mechanism to characterize the relationship between price dynamics and energy supply, and employ the SAA technique to approximate the risk-sensitive NE of the game. In doing so, they propose a distributed NE-seeking algorithm based on Nikaido-Isoda function and ADMM. The proposed method is validated through numerical simulations using real-world data from Australian energy market which shows that their algorithm can effectively reduce the risk of not meeting the demand and improve the economic benefits for each microgrid, compared to strategies that ignore uncertainty. Also, it is shown that a higher risk aversion weight reduces the energy bidding quantity, especially for renewable-based microgrids with high uncertainty.

Li et al. \cite{Li2016} study the problem of energy resource trading in an Energy Internet (EI) context. Similar to Li et al. \cite{li2017risk}, they consider a CVaR-sensitive two-stage MG with continuous state and action spaces, which includes a forward market and a spot market, to quantify the overbidding risk and uncertainties in energy supply and demand. The Cournot-based pricing mechanism is used to determine prices based on total expected/actual energy bids, while considering uncertainties in renewable generation and demand response in the spot market. The authors use an SAA technique to approximate the NE of the game and provide an NE-seeking algorithm based on the Nikaido-Isoda function and best response dynamics. They prove the existence of the approximated SAA NE under certain conditions. Finally, numerical simulations illustrate that the optimal risk-sensitive converged policies reduce overbidding risk compared to deterministic models that neglect uncertainty. Also, a higher risk price leads to more conservative bidding for microgrids with high uncertainty, e.g., wind, and more aggressive bidding for microgrids with low uncertainty, e.g., demand response. Furthermore, with decreasing deviation in the spot market capacities, microgrids tend to bid more energy to gain more benefits, coinciding with results from deterministic models.

Heidari et al. \cite{Heidari2022} explore how multi-energy carriers function as energy hubs within combined natural gas and electric power in competitive markets. They try to maximize profit in a joint-energy market using bi-level programming for strategic bidding and market clearing in both risk-neutral and risk-sensitive (CVaR) settings. The energy market MG model has a discrete state space and continuous action space and incorporates renewable energy uncertainty. Given the strict concavity of the payoff functions, the authors show the existence and uniqueness of NE and find it using a bi-level programming method. They show higher economic vulnerability of the system when the market operators are CVaR-sensitive compared to the risk-neutral setting.

Bolonhez et al. \cite{Bolonhez2022} consider decentralized blockchain financial networks, particularly Bitcoin. They propose a least-core-based quota allocation model for sharing mining pool rewards, considering mining uncertainty and risk aversion in a cooperative multiplayer MG with a discrete state space (coalitions) and a continuous action space (percentage of quotas allocated to coalition members). Numerical experiments show growing cooperative benefits over time but limitations due to dynamic reward changes. The authors also suggest potential improvements, including varied pool configurations and time-dependent mining probabilities to increase the "fairness" of their quota allocation method.

Li et al. \cite{Li2023} introduce a competitive risk-averse tri-level MG with defender and attacker players in supply networks. The model has a discrete state space, binary action space in levels one (defender's capacity backups) and two (attacker's facility attacks per locations), and a continous action space in level three (defender's recovered capacities at different facilities). The model is applied to supply networks to minimize the CvaR cost of the defender and protect against worst-case attacks. The article proposes defender recovery strategies while considering uncertainties in facility capacity and impact of attacks to enhance the resilience of supply networks against disruptions. Computational results show the effect of budget limitations and backup levels on optimal risk-sensitive policies, and that higher recovery ability reduces risk sensitivity of the defender.

Zhu et al. \cite{Zhu2022} examine a combined peer-to-peer (P2P) electricity market and carbon emission auction market modeled as an MG to promote localized energy trading and emission reduction for microgrids when the agents incorporate the CVaR risk criterion. They experimentally find the NE of the proposed MG using a multi-agent deep deterministic policy gradient (MADDPG) algorithm in a continuous state-action setting.

Lin et al. \cite{Lin2021} propose a p2p virtual power plant market bidding model for risk-aware energy trading among small-scale generators and consumers. The proposed model is a two-stage MG with continuous action and state spaces. The authors find the CVaR-sensitive Cournot NE of the game using a distributed ADMM algorithm. The equilibrium balances supply and demand of risk-sensitive market players. By analyzing case studies from the Australian energy market, the authors show that the proposed market structure can effectively reduce the overbidding risk while maximizing the renewable energy usage.

Munir et al. \cite{Munir20213476} consider energy challenges in IoT-supporting multi-access edge computing (MEC) networks. They introduce a risk-aware energy optimization model for MEC networks as an MG between energy suppliers and consumers in the network with continuous state and action spaces. The authors utilize asynchronous advantage actor-critic (A3C) \cite{mnih2016asynchronous} to find the CVaR-sensitive NE of the game.

Qiu et al. \cite{qiu2021rmix} propose RMIX, a cooperative value-based risk-sensitive algorithm for distributional MARL with continuous state and action spaces. In distributional RL, instead of learning the expected return, i.e., the mean of the return distribution, the aim is to learn the full distribution of returns. By learning the return distribution, distributional RL methods can capture the inherent randomness and risk in the environment more effectively. In these methods, instead of defining Q as the return of a given state-action pair $(s,a)$, we define it as a distribution $Z$ with respect to three r.v.s (reward, next state-action pair and its corresponding random return) as

\begin{equation}
    Z(s, a) \triangleq R(s, a)+\gamma Z\left(s^{\prime}, a^{\prime}\right).
\end{equation}

See Bellemare et al. \cite{bellemare2017distributional} for more details on distributional RL. The RMIX algorithm considers CVaR risks for agents over the learned distribution of $Z$ values in the above definition for each agent in a distributional cooperative MARL setting. The framework operates under centralized training with decentralized execution. It first learns each agent's return distribution to analytically calculate CVaR for decentralized execution. Afterwards, the algorithm includes a dynamic agent-specific risk level predictor to adjust agents' risk levels in real-time to handle the temporal nature of the stochastic outcomes during execution. Finally, CVaR policies are optimized by considering CVaR values as auxiliary local rewards and targets in the temporal difference (TD) error during centralized training, and updating local return distributions by minimizing a quantile regression loss function. The RMIX algorithm achieves state-of-the-art performance on cooperative cliff navigation scenarios and the challenging StarCraft II benchmark. An algorithm that is related to RMIX yet does not consider a traditional risk measure was proposed by Son et al. \cite{Son202220347}. The authors propose disentangled risk-sensitive multi-agent reinforcement learning (DRIMA), a distributional MARL framework based on IQN \cite{dabney2018implicit}, which disentangles the uncertainties arising from environment transitions and other agents' policies. They show that in a cooperative MARL settings, if agents act in a risk-neutral manner with respect to environmental uncertainties and "optimistically" with respect to other agents' policies, they outperform other state-of-the-art MARL methods, both distributional and non-distributional, in the challenging benchmark of StarCraft. The risk measure used in this work can be called "cooperative optimism", i.e., the agents focus on the higher quantiles of the action-value distribution and essentially assume that their teammates will take actions that lead to better outcomes, even if those actions have not been observed frequently during training.

Lyu et al. \cite{Lyu2020798} focus on decentralized cooperative MARL in partially observable MGs with continuous state and discrete action spaces, and with an optional risk-seeking setup, where the risk to be sought is equal to 1-CVaR, which they call conditional value-not-at-risk or CVnaR. The authors introduce a decentralized quantile estimator, based on IQN \cite{dabney2018implicit}, called likelihood quantile network (LQN). In this framework, each agent uses a decentralized quantile estimator to distinguish non-stationary samples based on the likelihood of returns. This helps in identifying samples that are influenced by other agents' exploration or sub-optimal policies. The experimental results show that risk-seeking behavior in cooperative MARL encourages exploration of high-reward spaces and converges faster compared to vanilla LQN.

Finally, Wang et al. \cite{Wang202222999} study risk-averse online convex competitive MGs with continuous state and action spaces, where agents seek to minimize a CVaR of their cost functions. The authors propose an online algorithm that approximates CVaR using zeroth-order gradient estimates and bandit feedback and achieves sub-linear regret with a high probability. They also introduce two enhancements for improving the regret bound, i.e., reusing samples from the previous iteration and residual feedback.

\subsubsection{Chance Constraint}

Zhong et al. \cite{Zhong2023440} consider multi-robot planning in a stochastic dynamic game framework with continuous state-action spaces and chance constraint risk based on Schwarting et al.'s game model \cite{schwarting2021stochastic} to capture agents' interactions and safety concerns. The paper presents the chance-constrained iterative linear-quadratic Markov games (CCILQGames) algorithm which uses augmented Lagrangian with automatic weight tuning to find the risk-sensitive NE of the game. They evaluate this algorithm in three autonomous driving scenarios of lane merging, intersection, and roundabout.

Yadollahi et al. \cite{yadollahi2023generalized} propose a generalized stochastic dynamic game with continuous state and action spaces and chance constraint risk criterion for modeling and analyzing demand-side management, in a microgrid where agents utilize both grid energy and a shared battery charged by renewable energy sources. The term \emph{generalized} in the context of demand-side management implies that there are local constraints for individual agents and shared coupling constraints (imposed on state of charge of the shared battery in the article) between agents in the game. The authors show the uniqueness of the risk-sensitive generalized NE, and provide an iterative algorithm to approximate this equilibrium. Simulation results demonstrate that the proposed stochastic model outperforms deterministic approaches with more effective peak shaving in the power exchange profile compared to the deterministic model.

\section{Trends and Future Directions}
\label{sec:discussion}

\begin{figure}[h]
\label{fig:year}
\begin{center}
\includegraphics[width=.99\linewidth]{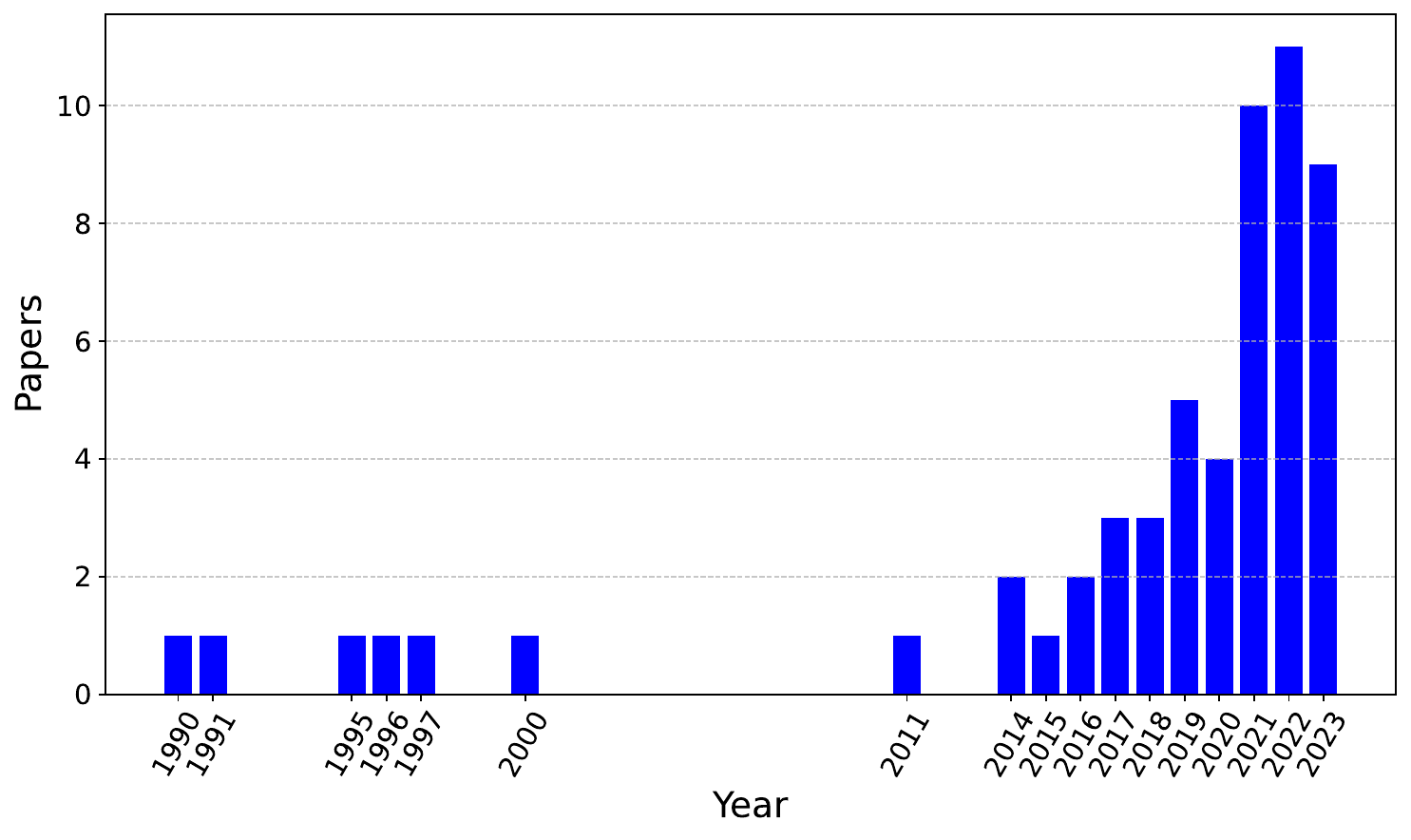}
\end{center}
\caption{Number of articles on risk-sensitive MG and MARL over the years}
\end{figure}

To observe the current trends and possible future directions of research in risk-sensitive MG and MARL, we extract the main characteristics of the reviewed studies, including risk measure, year, type of MG, and type of state and action spaces and provide this summary sorted by risk measure and year in Table~\ref{tab:main}. We also plot the number of articles per year in Figure~\ref{fig:year}. From these summaries, we observe that prior to 2016, the literature on risk-sensitive MG and MARL has been limited to the exponential reward/cost risk measure, and as we saw in Section~\ref{sec:res} almost exclusively to theoretical analysis. Although the trend of theoretical research for exponential risk has continued with a steady increase in the number of articles, other risk measures with motivations that are more application-oriented have become popular in recent years. This trend can be attributed to first, the need for specialized measures of risk for different application domains, and two, to the emergence of deep RL methods \cite{mnih2013playing,lillicrap2015continuous,silver2017mastering} since mid 2010s and availability of high-end GPUs that have enabled researchers to develop risk-sensitive multi-agent deep RL algorithms with different measures of risk akin to their specific application. With the increasing need for modeling and learning risk-sensitive behavior and policies in multi-agent systems in real-world scenarios, specifically in finance, energy trade, and autonomous driving, and with the rapid advancement of deep RL, we predict that the use of more diverse measures of risk, best fit for specific application domains continues to rise in the coming years.

\begin{sidewaystable*}[p]
\caption{Main characteristics of the articles reviewed. Expansion of short forms used: expo.: exponential reward/cost, np: N-player, zs: zero-sum, comp.: competitive, coop.: cooperative, coord.: coordination, cont.: continuous, dis.: discrete}
\label{tab:main}
\small
\begin{center}
\begin{tabular}{c|c|c|c|c||c|c|c|c|c}
\hline
Risk & Year & Study & Setup & State/Action & Risk & Year & Study & Setup & State/Action \\ \hline
& 2024 & \cite{Ghosh2024risk} & 2p zs & countable/Borel & & 1996 & \cite{Hernández-Hernández1996147} & 2p comp. & denumerable/Borel\\
& 2023 & \cite{Ghosh202340} & 2p zs & denumerable/Borel & Expo. & 1995 & \cite{Klompstra19952458} & 2p comp. & cont./cont.\\
& 2023 & \cite{baier2023} & 2p comp. & dis./dis. (both finite) & & 1991 & \cite{Krajewski1991260} & 2p comp. & cont./cont.\\
& 2023 & \cite{Bhabak2023134} & 2p zs \& non-zs & dis.-finite/Borel & & 1990 & \cite{Caravani1990117} & 2p comp. & cont./cont.\\ \cline{6-10}
& 2023 & \cite{Guo2023} & 2p zs & Borel/Borel & & 2023 & \cite{Li2023} & 2p comp. & dis./cont.\&dis.\\
& 2023 & \cite{Xu2023} & np comp. & cont./cont. & & 2022 & \cite{Bolonhez2022} & np coord. & dis./cont. \\
& 2022 & \cite{Golui202278} & 2p zs & denumerable/Borel & & 2022 & \cite{Zhu2022} & np comp. & cont./cont. \\
& 2022 & \cite{Golui2022485} & 2p zs & dis./Borel & & 2022 & \cite{Heidari2022} & np comp. & dis./cont. \\
& 2022 & \cite{ghosh2022nonzero} & 2p non-zsmg & denumerable/Borel  & & 2022 & \cite{Wang202222999} & np comp. & cont./cont.\\
& 2022 & \cite{Noorani20222266} & 2p coord. & dis./dis. & &  2021 & \cite{Lin2021} & np comp. & cont./cont.\\
& 2022 & \cite{wei2022risk} & np non-zs & denumerable/Borel & & 2021 & \cite{Munir20213476} & np comp. & cont./cont.\\
& 2022 & \cite{Pal202213} & 2p zs & Borel/cont. & CVaR & 2021 & \cite{qiu2021rmix} & np coop. & cont./cont.\\
& 2021 & \cite{Wei2021835} & 2p non-zs & denumerable/Borel & & 2021 & \cite{He20211398} & np comp. & cont./cont.\\
& 2021 & \cite{bacsar2021robust} & 3p zs and np mean-field& cont./cont. & & 2020 & \cite{Lyu2020798} & np coord. & cont./dis.\\
Expo. & 2021 & \cite{NaderiSoorki20217444} & np coop. & cont./dis. & & 2020 & \cite{Cui2020828} & np comp. & cont./cont.\\
& 2020 & \cite{Wang20206998} & 2p comp. & cont./cont. & & 2018 & \cite{Li20181386} & np comp. & cont./cont.\\
& 2019 & \cite{Asienkiewicz2019502} & 2p comp. & cont./cont. & & 2017 & \cite{li2017risk} & np comp.  & cont./cont.\\
& 2019 & \cite{Wei2019risk} & np coord. & denumerable/Borel & & 2016 & \cite{Li2016} & np comp. &  cont./cont.\\ \cline{6-10}
& 2019 & \cite{Cavazos-Cadena2019219} & 2p zsmg & dis./compact & & 2024 & \cite{ghaemi2024risk} & np comp. & dis./dis. (both finite)\\
& 2018 & \cite{Basu2018516} & 2p non-zs & countable/cont. & CPT & 2023 & \cite{shen2023riskq} & np. coop. & dis./dis. (both finite) \\
& 2017 & \cite{bauerle2022distributionally} & 2p zs & Borel/Borel & & 2021 & \cite{Tian20216011} & 2p comp. & dis./dis. (both finite)\\ 
& 2016 & \cite{Ghosh2016835} & 2p zs & denumerable/cont. & & 2018 & \cite{Xiao20182512} & 2p comp. & dis./dis. (both finite) \\ \cline{6-10}
& 2015 & \cite{Balbus2015247} & np bequest & cont./cont. &  & 2021 & \cite{FrédéricBonnans2021} & np mean-field & cont./cont.\\
& 2014 & \cite{basu2014zero} & 2p zs & countable/cont. & Coherent & 2020 & \cite{Huang20202022} & np comp. & dis./dis. (both finite)\\
& 2014 & \cite{Jaśkiewicz2014411} & np bequest & Borel/Borel & & 2017 & \cite{Pang2017235} & np comp.  & cont./cont.\\ \cline{6-10}
& 2011 & \cite{Tembine20114264} & np mean-field & dis.-finite/compact & Variance & 2021 & \cite{Parilina20211} & np coop. & dis./dis.\\
& 2000 & \cite{Klompstra2000nash} & 2p non-zs & cont./cont. & &2019 & \cite{Reddy20192171} & np mixed coop.-comp.  & cont./dis.\\ \cline{6-10}
& 1997 & \cite{Monahan1997149} & np comp. &  cont./cont. & Chance & 2023 & \cite{yadollahi2023generalized} & np comp. & cont./cont.\\ 
& & & & & & 2023 & \cite{Zhong2023440} & np comp. & cont./cont.\\ \cline{6-10}
& & & & & Coop. Optimism & 2022 & \cite{Son202220347} & np coord. &  cont./dis.
\end{tabular}
\end{center}
\end{sidewaystable*}

\section{Conclusion}
\label{sec:con}

In this work, we conducted a systematic review of risk sensitivity in Markov games and multi-agent reinforcement learning. We identified the risk measures used in the literature and defined and described them mathematically. Afterwards, we provided an in-depth analysis of the existing literature by considering each risk measure individually and discussing their related articles. Finally, we identified trends and future directions of research. We hope that this review facilitates and promotes additional research efforts in risk-sensitive multi-agent systems.




\bibliographystyle{IEEEtran}
\bibliography{refs}

\begin{thebibliography}{100}
\providecommand{\url}[1]{#1}
\csname url@samestyle\endcsname
\providecommand{\newblock}{\relax}
\providecommand{\bibinfo}[2]{#2}
\providecommand{\BIBentrySTDinterwordspacing}{\spaceskip=0pt\relax}
\providecommand{\BIBentryALTinterwordstretchfactor}{4}
\providecommand{\BIBentryALTinterwordspacing}{\spaceskip=\fontdimen2\font plus
\BIBentryALTinterwordstretchfactor\fontdimen3\font minus \fontdimen4\font\relax}
\providecommand{\BIBforeignlanguage}[2]{{%
\expandafter\ifx\csname l@#1\endcsname\relax
\typeout{** WARNING: IEEEtran.bst: No hyphenation pattern has been}%
\typeout{** loaded for the language `#1'. Using the pattern for}%
\typeout{** the default language instead.}%
\else
\language=\csname l@#1\endcsname
\fi
#2}}
\providecommand{\BIBdecl}{\relax}
\BIBdecl

\bibitem{shapley1953stochastic}
L.~S. Shapley, ``Stochastic games,'' \emph{Proceedings of the national academy of sciences}, vol.~39, no.~10, pp. 1095--1100, 1953.

\bibitem{littman1994markov}
M.~L. Littman, ``Markov games as a framework for multi-agent reinforcement learning,'' in \emph{Machine learning proceedings 1994}.\hskip 1em plus 0.5em minus 0.4em\relax Elsevier, 1994, pp. 157--163.

\bibitem{sayin2021decentralized}
M.~Sayin, K.~Zhang, D.~Leslie, T.~Basar, and A.~Ozdaglar, ``Decentralized q-learning in zero-sum markov games,'' \emph{Advances in Neural Information Processing Systems}, vol.~34, pp. 18\,320--18\,334, 2021.

\bibitem{zhang2020model}
K.~Zhang, S.~Kakade, T.~Basar, and L.~Yang, ``Model-based multi-agent rl in zero-sum markov games with near-optimal sample complexity,'' \emph{Advances in Neural Information Processing Systems}, vol.~33, pp. 1166--1178, 2020.

\bibitem{alacaoglu2022natural}
A.~Alacaoglu, L.~Viano, N.~He, and V.~Cevher, ``A natural actor-critic framework for zero-sum markov games,'' in \emph{International Conference on Machine Learning}.\hskip 1em plus 0.5em minus 0.4em\relax PMLR, 2022, pp. 307--366.

\bibitem{perolat2015approximate}
J.~Perolat, B.~Scherrer, B.~Piot, and O.~Pietquin, ``Approximate dynamic programming for two-player zero-sum markov games,'' in \emph{International Conference on Machine Learning}.\hskip 1em plus 0.5em minus 0.4em\relax PMLR, 2015, pp. 1321--1329.

\bibitem{qiu2021provably}
S.~Qiu, X.~Wei, J.~Ye, Z.~Wang, and Z.~Yang, ``Provably efficient fictitious play policy optimization for zero-sum markov games with structured transitions,'' in \emph{International Conference on Machine Learning}.\hskip 1em plus 0.5em minus 0.4em\relax PMLR, 2021, pp. 8715--8725.

\bibitem{chen2024finite}
Z.~Chen, K.~Zhang, E.~Mazumdar, A.~Ozdaglar, and A.~Wierman, ``A finite-sample analysis of payoff-based independent learning in zero-sum stochastic games,'' \emph{Advances in Neural Information Processing Systems}, vol.~36, 2024.

\bibitem{park2024multi}
C.~Park, K.~Zhang, and A.~Ozdaglar, ``Multi-player zero-sum markov games with networked separable interactions,'' \emph{Advances in Neural Information Processing Systems}, vol.~36, 2024.

\bibitem{ding2022independent}
D.~Ding, C.-Y. Wei, K.~Zhang, and M.~Jovanovic, ``Independent policy gradient for large-scale markov potential games: Sharper rates, function approximation, and game-agnostic convergence,'' in \emph{International Conference on Machine Learning}.\hskip 1em plus 0.5em minus 0.4em\relax PMLR, 2022, pp. 5166--5220.

\bibitem{fox2022independent}
R.~Fox, S.~M. Mcaleer, W.~Overman, and I.~Panageas, ``Independent natural policy gradient always converges in markov potential games,'' in \emph{International Conference on Artificial Intelligence and Statistics}.\hskip 1em plus 0.5em minus 0.4em\relax PMLR, 2022, pp. 4414--4425.

\bibitem{mguni2021learning}
D.~H. Mguni, Y.~Wu, Y.~Du, Y.~Yang, Z.~Wang, M.~Li, Y.~Wen, J.~Jennings, and J.~Wang, ``Learning in nonzero-sum stochastic games with potentials,'' in \emph{International Conference on Machine Learning}.\hskip 1em plus 0.5em minus 0.4em\relax PMLR, 2021, pp. 7688--7699.

\bibitem{leonardos2021global}
S.~Leonardos, W.~Overman, I.~Panageas, and G.~Piliouras, ``Global convergence of multi-agent policy gradient in markov potential games,'' \emph{arXiv preprint arXiv:2106.01969}, 2021.

\bibitem{maheshwari2022independent}
C.~Maheshwari, M.~Wu, D.~Pai, and S.~Sastry, ``Independent and decentralized learning in markov potential games,'' \emph{arXiv preprint arXiv:2205.14590}, 2022.

\bibitem{zhang2021multi}
K.~Zhang, Z.~Yang, and T.~Ba{\c{s}}ar, ``Multi-agent reinforcement learning: A selective overview of theories and algorithms,'' \emph{Handbook of reinforcement learning and control}, pp. 321--384, 2021.

\bibitem{canese2021multi}
L.~Canese, G.~C. Cardarilli, L.~Di~Nunzio, R.~Fazzolari, D.~Giardino, M.~Re, and S.~Span{\`o}, ``Multi-agent reinforcement learning: A review of challenges and applications,'' \emph{Applied Sciences}, vol.~11, no.~11, p. 4948, 2021.

\bibitem{yang2020overview}
Y.~Yang and J.~Wang, ``An overview of multi-agent reinforcement learning from game theoretical perspective,'' \emph{arXiv preprint arXiv:2011.00583}, 2020.

\bibitem{busoniu2008comprehensive}
L.~Busoniu, R.~Babuska, and B.~De~Schutter, ``A comprehensive survey of multiagent reinforcement learning,'' \emph{IEEE Transactions on Systems, Man, and Cybernetics, Part C (Applications and Reviews)}, vol.~38, no.~2, pp. 156--172, 2008.

\bibitem{nguyen2020deep}
T.~T. Nguyen, N.~D. Nguyen, and S.~Nahavandi, ``Deep reinforcement learning for multiagent systems: A review of challenges, solutions, and applications,'' \emph{IEEE transactions on cybernetics}, vol.~50, no.~9, pp. 3826--3839, 2020.

\bibitem{gronauer2022multi}
S.~Gronauer and K.~Diepold, ``Multi-agent deep reinforcement learning: a survey,'' \emph{Artificial Intelligence Review}, vol.~55, no.~2, pp. 895--943, 2022.

\bibitem{prashanth2022risk}
L.~Prashanth, M.~C. Fu \emph{et~al.}, ``Risk-sensitive reinforcement learning via policy gradient search,'' \emph{Foundations and Trends{\textregistered} in Machine Learning}, vol.~15, no.~5, pp. 537--693, 2022.

\bibitem{howard1972risk}
R.~A. Howard and J.~E. Matheson, ``Risk-sensitive markov decision processes,'' \emph{Management science}, vol.~18, no.~7, pp. 356--369, 1972.

\bibitem{artzner1999coherent}
P.~Artzner, F.~Delbaen, J.-M. Eber, and D.~Heath, ``Coherent measures of risk,'' \emph{Mathematical finance}, vol.~9, no.~3, pp. 203--228, 1999.

\bibitem{rockafellar2007coherent}
R.~T. Rockafellar, ``Coherent approaches to risk in optimization under uncertainty,'' in \emph{OR Tools and Applications: Glimpses of Future Technologies}.\hskip 1em plus 0.5em minus 0.4em\relax Informs, 2007, pp. 38--61.

\bibitem{kahneman1979prospect}
D.~Kahneman and A.~Tversky, ``Prospect theory: An analysis of decision under risk,'' \emph{Econometrica}, vol.~47, no.~2, pp. 363--391, 1979.

\bibitem{tversky1992advances}
A.~Tversky and D.~Kahneman, ``Advances in prospect theory: Cumulative representation of uncertainty,'' \emph{Journal of Risk and uncertainty}, vol.~5, pp. 297--323, 1992.

\bibitem{jie2018stochastic}
C.~Jie, L.~Prashanth, M.~Fu, S.~Marcus, and C.~Szepesv{\'a}ri, ``Stochastic optimization in a cumulative prospect theory framework,'' \emph{IEEE Transactions on Automatic Control}, vol.~63, no.~9, pp. 2867--2882, 2018.

\bibitem{denneberg1990distorted}
D.~Denneberg, ``Distorted probabilities and insurance premiums,'' \emph{Methods of Operations Research}, vol.~63, no.~3, pp. 3--5, 1990.

\bibitem{sereda2010distortion}
E.~N. Sereda, E.~M. Bronshtein, S.~T. Rachev, F.~J. Fabozzi, W.~Sun, and S.~V. Stoyanov, ``Distortion risk measures in portfolio optimization,'' \emph{Handbook of portfolio construction}, pp. 649--673, 2010.

\bibitem{vijayan2021policy}
N.~Vijayan \emph{et~al.}, ``Policy gradient methods for distortion risk measures,'' \emph{arXiv preprint arXiv:2107.04422}, 2021.

\bibitem{sobel1982variance}
M.~J. Sobel, ``The variance of discounted markov decision processes,'' \emph{Journal of Applied Probability}, vol.~19, no.~4, pp. 794--802, 1982.

\bibitem{filar1989variance}
J.~A. Filar, L.~C. Kallenberg, and H.-M. Lee, ``Variance-penalized markov decision processes,'' \emph{Mathematics of Operations Research}, vol.~14, no.~1, pp. 147--161, 1989.

\bibitem{rockafellar2000optimization}
R.~T. Rockafellar, S.~Uryasev \emph{et~al.}, ``Optimization of conditional value-at-risk,'' \emph{Journal of risk}, vol.~2, pp. 21--42, 2000.

\bibitem{ang2018dual}
M.~Ang, J.~Sun, and Q.~Yao, ``On the dual representation of coherent risk measures,'' \emph{Annals of Operations Research}, vol. 262, pp. 29--46, 2018.

\bibitem{charnes1958cost}
A.~Charnes, W.~W. Cooper, and G.~H. Symonds, ``Cost horizons and certainty equivalents: an approach to stochastic programming of heating oil,'' \emph{Management science}, vol.~4, no.~3, pp. 235--263, 1958.

\bibitem{page2021prisma}
M.~J. Page, J.~E. McKenzie, P.~M. Bossuyt, I.~Boutron, T.~C. Hoffmann, C.~D. Mulrow, L.~Shamseer, J.~M. Tetzlaff, E.~A. Akl, S.~E. Brennan \emph{et~al.}, ``The prisma 2020 statement: an updated guideline for reporting systematic reviews,'' \emph{Bmj}, vol. 372, 2021.

\bibitem{Başar1999479}
T.~Başar, ``\BIBforeignlanguage{English}{Nash equilibria of risk-sensitive nonlinear stochastic differential games},'' \emph{\BIBforeignlanguage{English}{Journal of Optimization Theory and Applications}}, vol. 100, no.~3, pp. 479--498, 1999.

\bibitem{Golui2023327}
S.~Golui and C.~Pal, ``Continucontinuous-time zero-sum games for markov decision processes with discounted risk-sensitive cost criterion on a general state spaceous-time zero-sum games for markov decision processes with discounted risk-sensitive cost criterion on a general state space,'' \emph{Stochastic Analysis and Applications}, vol.~41, no.~2, p. 327 – 357, 2023.

\bibitem{zhang2021provably}
Y.~Zhang, Z.~Yang, and Z.~Wang, ``Provably efficient actor-critic for risk-sensitive and robust adversarial rl: A linear-quadratic case,'' in \emph{International Conference on Artificial Intelligence and Statistics}.\hskip 1em plus 0.5em minus 0.4em\relax PMLR, 2021, pp. 2764--2772.

\bibitem{Tang20031065}
C.~Tang and T.~Başar, ``\BIBforeignlanguage{English}{Minimax nonlinear control under stochastic uncertainty constraints},'' vol.~1, 2003, pp. 1065--1070.

\bibitem{Dvijotham2011179}
K.~Dvijotham and E.~Todorov, ``\BIBforeignlanguage{English}{A unifying framework for linearly solvable control}.''\hskip 1em plus 0.5em minus 0.4em\relax AUAI Press, 2011, pp. 179--186.

\bibitem{zhang2021derivative}
K.~Zhang, X.~Zhang, B.~Hu, and T.~Basar, ``Derivative-free policy optimization for linear risk-sensitive and robust control design: Implicit regularization and sample complexity,'' \emph{Advances in Neural Information Processing Systems}, vol.~34, pp. 2949--2964, 2021.

\bibitem{Dey19971587}
S.~Dey and J.~Moore, ``\BIBforeignlanguage{English}{Risk-sensitive filtering and smoothing via reference probability methods},'' \emph{\BIBforeignlanguage{English}{IEEE Transactions on Automatic Control}}, vol.~42, no.~11, pp. 1587--1591, 1997.

\bibitem{Fleming19971790}
W.~Fleming and D.~Hernández-Hernández, ``\BIBforeignlanguage{English}{Risk-sensitive control of finite state machines on an infinite horizon i},'' \emph{\BIBforeignlanguage{English}{SIAM Journal on Control and Optimization}}, vol.~35, no.~5, pp. 1790--1810, 1997.

\bibitem{Ding202239}
R.~Ding and E.~Feinberg, ``\BIBforeignlanguage{English}{Cvar optimization for mdps: Existence and computation of optimal policies},'' \emph{\BIBforeignlanguage{English}{Performance Evaluation Review}}, vol.~50, no.~2, pp. 39--41, 2022.

\bibitem{Ghosh202340}
M.~Ghosh, S.~Golui, C.~Pal, and S.~Pradhan, ``\BIBforeignlanguage{English}{Discrete-time zero-sum games for markov chains with risk-sensitive average cost criterion},'' \emph{\BIBforeignlanguage{English}{Stochastic Processes and their Applications}}, vol. 158, pp. 40--74, 2023.

\bibitem{Wei2021835}
Q.~Wei and X.~Chen, ``\BIBforeignlanguage{English}{Nonzero-sum risk-sensitive average stochastic games: The case of unbounded costs},'' \emph{\BIBforeignlanguage{English}{Dynamic Games and Applications}}, vol.~11, no.~4, pp. 835--862, 2021.

\bibitem{Wei2019risk}
------, ``Risk-sensitive average equilibria for discrete-time stochastic games,'' \emph{Dynamic Games and Applications}, vol.~9, pp. 521--549, 2019.

\bibitem{Hernández-Hernández1996147}
D.~Hernández-Hernández and S.~Marcus, ``\BIBforeignlanguage{English}{Risk sensitive control of markov processes in countable state space},'' \emph{\BIBforeignlanguage{English}{Systems and Control Letters}}, vol.~29, no.~3, pp. 147--155, 1996.

\bibitem{Golui202278}
S.~Golui and C.~Pal, ``\BIBforeignlanguage{English}{Continuous-time zero-sum games for markov chains with risk-sensitive finite-horizon cost criterion},'' \emph{\BIBforeignlanguage{English}{Stochastic Analysis and Applications}}, vol.~40, no.~1, pp. 78--95, 2022.

\bibitem{Wei201996}
Q.~Wei, ``\BIBforeignlanguage{English}{Nonzero-sum risk-sensitive finite-horizon continuous-time stochastic games},'' \emph{\BIBforeignlanguage{English}{Statistics and Probability Letters}}, vol. 147, pp. 96--104, 2019.

\bibitem{ghosh2022nonzero}
M.~K. Ghosh, S.~Golui, C.~Pal, and S.~Pradhan, ``Nonzero-sum risk-sensitive continuous-time stochastic games with ergodic costs,'' \emph{Applied Mathematics \& Optimization}, vol.~86, no.~1, p.~6, 2022.

\bibitem{Caravani1990117}
P.~Caravani and G.~Papavassilopoulos, ``\BIBforeignlanguage{English}{A class of risk-sensitive noncooperative games},'' \emph{\BIBforeignlanguage{English}{Journal of Economic Dynamics and Control}}, vol.~14, no.~1, pp. 117--149, 1990.

\bibitem{Wang20206998}
M.~Wang, N.~Mehr, A.~Gaidon, and M.~Schwager, ``\BIBforeignlanguage{English}{Game-theoretic planning for risk-aware interactive agents}.''\hskip 1em plus 0.5em minus 0.4em\relax Institute of Electrical and Electronics Engineers Inc., 2020, pp. 6998--7005.

\bibitem{Krajewski1991260}
W.~Krajewski, ``\BIBforeignlanguage{English}{Note on the dynamic risk sensitive nash games},'' \emph{\BIBforeignlanguage{English}{Lecture Notes in Control and Information Sciences}}, vol. 157, pp. 260--268, 1991.

\bibitem{Jaśkiewicz2014411}
A.~Jaśkiewicz and A.~Nowak, ``\BIBforeignlanguage{English}{Stationary markov perfect equilibria in risk sensitive stochastic overlapping generations models},'' \emph{\BIBforeignlanguage{English}{Journal of Economic Theory}}, vol. 151, no.~1, pp. 411--447, 2014.

\bibitem{Balbus2015247}
T.~Balbus, A.~Jaśkiewicz, and A.~Nowak, ``\BIBforeignlanguage{English}{Stochastic bequest games},'' \emph{\BIBforeignlanguage{English}{Games and Economic Behavior}}, vol.~90, pp. 247--256, 2015.

\bibitem{Klompstra2000nash}
M.~B. Klompstra, ``Nash equilibria in risk-sensitive dynamic games,'' \emph{IEEE Transactions on Automatic Control}, vol.~45, no.~7, pp. 1397--1401, 2000.

\bibitem{Klompstra19952458}
M.~Klompstra, ``\BIBforeignlanguage{English}{Nash equilibria in risk-sensitive dynamic games},'' vol.~3, 1995, pp. 2458--2462.

\bibitem{Xu2023}
R.~Xu and T.~Wu, ``Risk-sensitive large-population linear-quadratic-gaussian games with major and minor agents,'' \emph{Asian Journal of Control}, 2023.

\bibitem{NaderiSoorki20217444}
M.~Naderi~Soorki, W.~Saad, M.~Bennis, and C.~Hong, ``\BIBforeignlanguage{English}{Ultra-reliable indoor millimeter wave communications using multiple artificial intelligence-powered intelligent surfaces},'' \emph{\BIBforeignlanguage{English}{IEEE Transactions on Communications}}, vol.~69, no.~11, pp. 7444--7457, 2021.

\bibitem{bacsar2021robust}
T.~Ba{\c{s}}ar, ``Robust designs through risk sensitivity: An overview,'' \emph{Journal of Systems Science and Complexity}, vol.~34, pp. 1634--1665, 2021.

\bibitem{Bhabak2023134}
A.~Bhabak and S.~Saha, ``\BIBforeignlanguage{English}{Zero and non-zero sum risk-sensitive semi-markov games},'' \emph{\BIBforeignlanguage{English}{Stochastic Analysis and Applications}}, vol.~41, no.~1, pp. 134--151, 2023.

\bibitem{Ghosh2024risk}
M.~K. Ghosh, S.~Golui, C.~Pal, and S.~Pradhan, ``Zero-sum stochastic games in continuous-time with risk-sensitive average cost crition on a countable state space,'' \emph{Mathematical Control and Related Fields}, 2024.

\bibitem{baier2023}
C.~Baier, K.~Chatterjee, T.~Meggendorfer, and J.~Piribauer, ``{Entropic Risk for Turn-Based Stochastic Games},'' in \emph{48th International Symposium on Mathematical Foundations of Computer Science (MFCS 2023)}, ser. Leibniz International Proceedings in Informatics (LIPIcs), J.~Leroux, S.~Lombardy, and D.~Peleg, Eds., vol. 272.\hskip 1em plus 0.5em minus 0.4em\relax Dagstuhl, Germany: Schloss Dagstuhl -- Leibniz-Zentrum f{\"u}r Informatik, 2023, pp. 15:1--15:16.

\bibitem{Golui2022485}
S.~Golui, C.~Pal, and S.~Saha, ``\BIBforeignlanguage{English}{Continuous-time zero-sum games for markov decision processes with discounted risk-sensitive cost criterion},'' \emph{\BIBforeignlanguage{English}{Dynamic Games and Applications}}, vol.~12, no.~2, pp. 485--512, 2022.

\bibitem{Pal202213}
C.~Pal and S.~Pradhan, ``\BIBforeignlanguage{English}{Zero-sum games for pure jump processes with risk-sensitive discounted cost criteria},'' \emph{\BIBforeignlanguage{English}{Journal of Dynamics and Games}}, vol.~9, no.~1, pp. 13--25, 2022.

\bibitem{Guo2023}
X.~Guo, J.~Chen, and Z.~Li, ``\BIBforeignlanguage{English}{Zero-sum risk-sensitive stochastic games with unbounded payoff functions and varying discount factors},'' \emph{\BIBforeignlanguage{English}{Journal of Mathematical Analysis and Applications}}, vol. 519, no.~2, 2023.

\bibitem{Asienkiewicz2019502}
H.~Asienkiewicz and L.~Balbus, ``\BIBforeignlanguage{English}{Existence of nash equilibria in stochastic games of resource extraction with risk-sensitive players},'' \emph{\BIBforeignlanguage{English}{TOP}}, vol.~27, no.~3, pp. 502--518, 2019.

\bibitem{Monahan1997149}
G.~Monahan and M.~Sobel, ``\BIBforeignlanguage{English}{Risk-sensitive dynamic market share attraction games},'' \emph{\BIBforeignlanguage{English}{Games and Economic Behavior}}, vol.~20, no.~2, pp. 149--160, 1997.

\bibitem{wei2022risk}
Q.~Wei and X.~Chen, ``Risk-sensitive first passage stochastic games with unbounded costs,'' \emph{Optimization}, pp. 1--34, 2022.

\bibitem{Noorani20222266}
E.~Noorani and J.~Baras, ``\BIBforeignlanguage{English}{Risk-attitudes, trust, and emergence of coordination in multi-agent reinforcement learning systems: A study of independent risk-sensitive reinforce}.''\hskip 1em plus 0.5em minus 0.4em\relax Institute of Electrical and Electronics Engineers Inc., 2022, pp. 2266--2271.

\bibitem{williams1992simple}
R.~J. Williams, ``Simple statistical gradient-following algorithms for connectionist reinforcement learning,'' \emph{Machine learning}, vol.~8, pp. 229--256, 1992.

\bibitem{Basu2018516}
A.~Basu and M.~Ghosh, ``\BIBforeignlanguage{English}{Nonzero-sum risk-sensitive stochastic games on a countable state space},'' \emph{\BIBforeignlanguage{English}{Mathematics of Operations Research}}, vol.~43, no.~2, pp. 516--532, 2018.

\bibitem{basu2014zero}
A.~Basu and M.~K. Ghosh, ``Zero-sum risk-sensitive stochastic games on a countable state space,'' \emph{Stochastic processes and their applications}, vol. 124, no.~1, pp. 961--983, 2014.

\bibitem{bauerle2022distributionally}
N.~B{\"a}uerle and A.~Glauner, ``Distributionally robust markov decision processes and their connection to risk measures,'' \emph{Mathematics of Operations Research}, vol.~47, no.~3, pp. 1757--1780, 2022.

\bibitem{Ghosh2016835}
M.~Ghosh, K.~Kumar, and C.~Pal, ``\BIBforeignlanguage{English}{Zero-sum risk-sensitive stochastic games for continuous time markov chains},'' \emph{\BIBforeignlanguage{English}{Stochastic Analysis and Applications}}, vol.~34, no.~5, pp. 835--851, 2016.

\bibitem{Cavazos-Cadena2019219}
R.~Cavazos-Cadena and D.~Hernández-Hernández, ``\BIBforeignlanguage{English}{The vanishing discount approach in a class of zero-sum finite games with risk-sensitive average criterion},'' \emph{\BIBforeignlanguage{English}{SIAM Journal on Control and Optimization}}, vol.~57, no.~1, pp. 219--240, 2019.

\bibitem{Tembine20114264}
H.~Tembine, ``\BIBforeignlanguage{English}{Risk-sensitive mean field stochastic games}.''\hskip 1em plus 0.5em minus 0.4em\relax Institute of Electrical and Electronics Engineers Inc., 2011, pp. 4264--4269.

\bibitem{FrédéricBonnans2021}
J.~Frédéric~Bonnans, P.~Lavigne, and L.~Pfeiffer, ``\BIBforeignlanguage{English}{Discrete-time mean field games with risk-averse agents},'' \emph{\BIBforeignlanguage{English}{ESAIM - Control, Optimisation and Calculus of Variations}}, vol.~27, 2021.

\bibitem{Pang2017235}
J.-S. Pang, S.~Sen, and U.~Shanbhag, ``\BIBforeignlanguage{English}{Two-stage non-cooperative games with risk-averse players},'' \emph{\BIBforeignlanguage{English}{Mathematical Programming}}, vol. 165, no.~1, pp. 235--290, 2017.

\bibitem{Huang20202022}
W.~Huang, P.~Hai, and W.~Haskell, ``\BIBforeignlanguage{English}{Model and reinforcement learning for markov games with risk preferences}.''\hskip 1em plus 0.5em minus 0.4em\relax AAAI press, 2020, pp. 2022--2029.

\bibitem{Tian20216011}
R.~Tian, L.~Sun, and M.~Tomizuka, ``\BIBforeignlanguage{English}{Bounded risk-sensitive markov games: Forward policy design and inverse reward learning with iterative reasoning and cumulative prospect theory},'' vol.~7.\hskip 1em plus 0.5em minus 0.4em\relax Association for the Advancement of Artificial Intelligence, 2021, pp. 6011--6020.

\bibitem{lin2018probabilistically}
K.~Lin, C.~Jie, and S.~I. Marcus, ``Probabilistically distorted risk-sensitive infinite-horizon dynamic programming,'' \emph{Automatica}, vol.~97, pp. 1--6, 2018.

\bibitem{wright2017predicting}
J.~R. Wright and K.~Leyton-Brown, ``Predicting human behavior in unrepeated, simultaneous-move games,'' \emph{Games and Economic Behavior}, vol. 106, pp. 16--37, 2017.

\bibitem{Xiao20182512}
L.~Xiao, D.~Xu, N.~Mandayam, and H.~Poor, ``\BIBforeignlanguage{English}{Attacker-centric view of a detection game against advanced persistent threats},'' \emph{\BIBforeignlanguage{English}{IEEE Transactions on Mobile Computing}}, vol.~17, no.~11, pp. 2512--2523, 2018.

\bibitem{shen2023riskq}
S.~Shen, C.~Ma, C.~Li, W.~Liu, Y.~Fu, S.~Mei, X.~Liu, and C.~Wang, ``Riskq: risk-sensitive multi-agent reinforcement learning value factorization,'' \emph{Advances in Neural Information Processing Systems}, vol.~36, pp. 34\,791--34\,825, 2023.

\bibitem{dabney2018implicit}
W.~Dabney, G.~Ostrovski, D.~Silver, and R.~Munos, ``Implicit quantile networks for distributional reinforcement learning,'' in \emph{International conference on machine learning}.\hskip 1em plus 0.5em minus 0.4em\relax PMLR, 2018, pp. 1096--1105.

\bibitem{ghaemi2024risk}
H.~Ghaemi, H.~Kebriaei, A.~Ramezani~Moghaddam, and M.~Nili~Ahmadabadi, ``Risk-sensitive multi-agent reinforcement learning in network aggregative markov games,'' in \emph{Proceedings of the 23rd International Conference on Autonomous Agents and Multiagent Systems}, ser. AAMAS '24.\hskip 1em plus 0.5em minus 0.4em\relax Richland, SC: International Foundation for Autonomous Agents and Multiagent Systems, 2024, p. 2282–2284.

\bibitem{moghaddam2024expected}
A.~R. Moghaddam and H.~Kebriaei, ``Expected policy gradient for network aggregative markov games in continuous space,'' \emph{IEEE Transactions on Neural Networks and Learning Systems}, 2024.

\bibitem{parise2020distributed}
F.~Parise, S.~Grammatico, B.~Gentile, and J.~Lygeros, ``Distributed convergence to nash equilibria in network and average aggregative games,'' \emph{Automatica}, vol. 117, p. 108959, 2020.

\bibitem{shokri2020leader}
M.~Shokri and H.~Kebriaei, ``Leader--follower network aggregative game with stochastic agents’ communication and activeness,'' \emph{IEEE Transactions on Automatic Control}, vol.~65, no.~12, pp. 5496--5502, 2020.

\bibitem{Reddy20192171}
D.~Reddy, A.~Saha, S.~Tamilselvam, P.~Agrawal, and P.~Dayama, ``\BIBforeignlanguage{English}{Risk averse reinforcement learning for mixed multi-agent environments},'' vol.~4.\hskip 1em plus 0.5em minus 0.4em\relax International Foundation for Autonomous Agents and Multiagent Systems (IFAAMAS), 2019, pp. 2171--2173.

\bibitem{lowe2017multi}
R.~Lowe, Y.~I. Wu, A.~Tamar, J.~Harb, O.~Pieter~Abbeel, and I.~Mordatch, ``Multi-agent actor-critic for mixed cooperative-competitive environments,'' \emph{Advances in neural information processing systems}, vol.~30, 2017.

\bibitem{Parilina20211}
E.~Parilina and S.~Akimochkin, ``\BIBforeignlanguage{English}{Cooperative stochastic games with mean-variance preferences},'' \emph{\BIBforeignlanguage{English}{Mathematics}}, vol.~9, no.~3, pp. 1--15, 2021.

\bibitem{suijs1999cooperative}
J.~Suijs, P.~Borm, A.~De~Waegenaere, and S.~Tijs, ``Cooperative games with stochastic payoffs,'' \emph{European Journal of Operational Research}, vol. 113, no.~1, pp. 193--205, 1999.

\bibitem{He20211398}
S.~He, S.~Yu, L.~Wang, Y.~Liu, X.~Lin, and W.~Han, ``\BIBforeignlanguage{English}{Trade-off stochastic game based bidding strategy for multiple wind farms}.''\hskip 1em plus 0.5em minus 0.4em\relax Institute of Electrical and Electronics Engineers Inc., 2021, pp. 1398--1403.

\bibitem{Cui2020828}
S.~Cui, Y.-W. Wang, C.~Li, and J.-W. Xiao, ``\BIBforeignlanguage{English}{Prosumer community: A risk aversion energy sharing model},'' \emph{\BIBforeignlanguage{English}{IEEE Transactions on Sustainable Energy}}, vol.~11, no.~2, pp. 828--838, 2020.

\bibitem{Li20181386}
C.~Li, C.~Liu, K.~Deng, X.~Yu, and T.~Huang, ``\BIBforeignlanguage{English}{Data-driven charging strategy of pevs under transformer aging risk},'' \emph{\BIBforeignlanguage{English}{IEEE Transactions on Control Systems Technology}}, vol.~26, no.~4, pp. 1386--1399, 2018.

\bibitem{wei2012distributed}
E.~Wei and A.~Ozdaglar, ``Distributed alternating direction method of multipliers,'' in \emph{2012 IEEE 51st IEEE Conference on Decision and Control (CDC)}.\hskip 1em plus 0.5em minus 0.4em\relax IEEE, 2012, pp. 5445--5450.

\bibitem{li2017risk}
C.~Li, Y.~Xu, X.~Yu, C.~Ryan, and T.~Huang, ``Risk-averse energy trading in multienergy microgrids: A two-stage stochastic game approach,'' \emph{IEEE Transactions on Industrial Informatics}, vol.~13, no.~5, pp. 2620--2630, 2017.

\bibitem{Li2016}
C.~Li, X.~Yu, P.~Sokolowski, N.~Liu, and G.~Chen, ``\BIBforeignlanguage{English}{A stochastic game for energy resource trading in the context of energy internet},'' vol. 2016-November.\hskip 1em plus 0.5em minus 0.4em\relax IEEE Computer Society, 2016.

\bibitem{Heidari2022}
A.~Heidari, R.~Bansal, J.~Hossain, and J.~Zhu, ``\BIBforeignlanguage{English}{Strategic risk aversion of smart energy hubs in the joined energy markets applying a stochastic game approach},'' \emph{\BIBforeignlanguage{English}{Journal of Cleaner Production}}, vol. 349, 2022.

\bibitem{Bolonhez2022}
E.~Bolonhez, T.~Silva, and B.~Fanzeres, ``\BIBforeignlanguage{English}{A core-based quota allocation model for the bitcoin-refunded blockchain network},'' \emph{\BIBforeignlanguage{English}{Expert Systems with Applications}}, vol. 209, 2022.

\bibitem{Li2023}
Q.~Li, M.~Li, Y.~Tian, and J.~Gan, ``\BIBforeignlanguage{English}{A risk-averse tri-level stochastic model for locating and recovering facilities against attacks in an uncertain environment},'' \emph{\BIBforeignlanguage{English}{Reliability Engineering and System Safety}}, vol. 229, 2023.

\bibitem{Zhu2022}
Z.~Zhu, K.~Chan, S.~Bu, B.~Zhou, and S.~Xia, ``\BIBforeignlanguage{English}{Nash equilibrium estimation and analysis in joint peer-to-peer electricity and carbon emission auction market with microgrid prosumers},'' \emph{\BIBforeignlanguage{English}{IEEE Transactions on Power Systems}}, pp. 1--13, 2022.

\bibitem{Lin2021}
W.-T. Lin, G.~Chen, and C.~Li, ``\BIBforeignlanguage{English}{Risk-averse energy trading among peer-to-peer based virtual power plants: A stochastic game approach},'' \emph{\BIBforeignlanguage{English}{International Journal of Electrical Power and Energy Systems}}, vol. 132, 2021.

\bibitem{Munir20213476}
M.~Munir, S.~Abedin, N.~Tran, Z.~Han, E.-N. Huh, and C.~Hong, ``\BIBforeignlanguage{English}{Risk-aware energy scheduling for edge computing with microgrid: A multi-agent deep reinforcement learning approach},'' \emph{\BIBforeignlanguage{English}{IEEE Transactions on Network and Service Management}}, vol.~18, no.~3, pp. 3476--3497, 2021.

\bibitem{mnih2016asynchronous}
V.~Mnih, A.~P. Badia, M.~Mirza, A.~Graves, T.~Lillicrap, T.~Harley, D.~Silver, and K.~Kavukcuoglu, ``Asynchronous methods for deep reinforcement learning,'' in \emph{International conference on machine learning}.\hskip 1em plus 0.5em minus 0.4em\relax PMLR, 2016, pp. 1928--1937.

\bibitem{qiu2021rmix}
W.~Qiu, X.~Wang, R.~Yu, R.~Wang, X.~He, B.~An, S.~Obraztsova, and Z.~Rabinovich, ``Rmix: Learning risk-sensitive policies for cooperative reinforcement learning agents,'' \emph{Advances in Neural Information Processing Systems}, vol.~34, pp. 23\,049--23\,062, 2021.

\bibitem{bellemare2017distributional}
M.~G. Bellemare, W.~Dabney, and R.~Munos, ``A distributional perspective on reinforcement learning,'' in \emph{International conference on machine learning}.\hskip 1em plus 0.5em minus 0.4em\relax PMLR, 2017, pp. 449--458.

\bibitem{Son202220347}
K.~Son, J.~Kim, S.~Ahn, R.~D. Reyes, Y.~Yi, and J.~Shin, ``Disentangling sources of risk for distributional multi-agent reinforcement learning,'' C.~K., J.~S., S.~L., S.~C., N.~G., and S.~S., Eds., vol. 162.\hskip 1em plus 0.5em minus 0.4em\relax ML Research Press, 2022, Conference paper, p. 20347 – 20368.

\bibitem{Lyu2020798}
X.~Lyu and C.~Amato, ``\BIBforeignlanguage{English}{Likelihood quantile networks for coordinating multi-agent reinforcement learning},'' S.~G. An~B., El Fallah Seghrouchni~A., Ed., vol. 2020-May.\hskip 1em plus 0.5em minus 0.4em\relax International Foundation for Autonomous Agents and Multiagent Systems (IFAAMAS), 2020, pp. 798--806.

\bibitem{Wang202222999}
Z.~Wang, Y.~Shen, and M.~M. Zavlanos, ``Risk-averse no-regret learning in online convex games,'' C.~K., J.~S., S.~L., S.~C., N.~G., and S.~S., Eds., vol. 162.\hskip 1em plus 0.5em minus 0.4em\relax ML Research Press, 2022, Conference paper, p. 22999 – 23017.

\bibitem{Zhong2023440}
H.~Zhong, Y.~Shimizu, and J.~Chen, ``\BIBforeignlanguage{English}{Chance-constrained iterative linear-quadratic stochastic games},'' \emph{\BIBforeignlanguage{English}{IEEE Robotics and Automation Letters}}, vol.~8, no.~1, pp. 440--447, 2023.

\bibitem{schwarting2021stochastic}
W.~Schwarting, A.~Pierson, S.~Karaman, and D.~Rus, ``Stochastic dynamic games in belief space,'' \emph{IEEE Transactions on Robotics}, vol.~37, no.~6, pp. 2157--2172, 2021.

\bibitem{yadollahi2023generalized}
S.~Yadollahi, H.~Kebriaei, and S.~Soudjani, ``Generalized stochastic dynamic aggregative game for demand-side management in microgrids with shared battery,'' \emph{IEEE Control Systems Letters}, 2023.

\bibitem{mnih2013playing}
V.~Mnih, K.~Kavukcuoglu, D.~Silver, A.~Graves, I.~Antonoglou, D.~Wierstra, and M.~Riedmiller, ``Playing atari with deep reinforcement learning,'' \emph{arXiv preprint arXiv:1312.5602}, 2013.

\bibitem{lillicrap2015continuous}
T.~P. Lillicrap, J.~J. Hunt, A.~Pritzel, N.~Heess, T.~Erez, Y.~Tassa, D.~Silver, and D.~Wierstra, ``Continuous control with deep reinforcement learning,'' \emph{arXiv preprint arXiv:1509.02971}, 2015.

\bibitem{silver2017mastering}
D.~Silver, J.~Schrittwieser, K.~Simonyan, I.~Antonoglou, A.~Huang, A.~Guez, T.~Hubert, L.~Baker, M.~Lai, A.~Bolton \emph{et~al.}, ``Mastering the game of go without human knowledge,'' \emph{nature}, vol. 550, no. 7676, pp. 354--359, 2017.

\end{thebibliography}

\end{document}